\documentclass{article}
\usepackage[utf8]{inputenc}
\usepackage[T1]{fontenc} 
\usepackage[a4paper]{geometry}
\usepackage[english]{babel}		   
\usepackage{amsmath,amssymb, amsthm} 
\usepackage[font=small, labelfont=bf, labelsep=period]{caption}
\usepackage{graphicx}
\usepackage{url}
\usepackage{hyperref}
\usepackage{xcolor}
\usepackage{mathrsfs}
\usepackage{authblk}
\usepackage{breakcites}
\usepackage{natbib}
\usepackage[flushleft]{threeparttable}
\usepackage{lscape}

\newcommand{\R}{\mathbb{R}}                  
\newcommand{\T}{\mathbb{T}}

\DeclareMathOperator{\CHI}{\mathcal{X}}
\def\n{\mathtt n}
\def\k{\mathtt{k}}
\def\H{\mathcal H}
\def\G{\mathcal G}
\def\A{\mathcal{A}}
\def\calA{\mathcal{A}}

\def\calE{\mathcal{E}}
\def\calW{\mathcal{W}}

\numberwithin{equation}{section}

\title{Proper elements for resonant planet-crossing asteroids}

\author[1]{M. Fenucci\thanks{email: \texttt{marco\_fenucci@matf.bg.ac.rs}}}
\author[2]{G. F. Gronchi}
\author[3]{M. Saillenfest}

\affil[1]{Department of Astronomy, Faculty of Mathematics, University
  of Belgrade, Studentski trg 16, 11000 Belgrade, Serbia}
\affil[2]{Dipartimento di Matematica, Università di Pisa, Largo
  B. Pontecorvo 5, 56127 Pisa, Italy}
\affil[3]{IMCCE Observatoire de Paris, PSL Research University, CNRS, Sorbonne Université, Université de Lille, 75014 Paris, France}

\date{\today}

\begin{document}

\maketitle

\begin{abstract}
   Proper elements are quasi-integrals of motion of a dynamical system, 
   meaning that they can be considered constant over a 
   certain timespan, and they permit to describe the long-term evolution of the
   system with a few
   parameters.
   Near-Earth objects (NEOs) generally have a large eccentricity and therefore they
   can cross the orbits of the planets. Moreover, some of them are known to be currently in a
   mean-motion resonance with a planet. Thus, the methods previously used for the
   computation of main-belt asteroid proper elements are not appropriate for such objects. 

   In this paper, we introduce a technique for the computation of proper elements of
   planet-crossing asteroids that are in a mean motion resonance with a planet.
   First, we numerically average the Hamiltonian over the fast angles while keeping all the resonant
   terms, and we describe how to continue a solution beyond orbit crossing singularities. 
   Proper elements are then extracted by a frequency analysis of the averaged orbit-crossing solutions.
   We give proper elements of some known resonant NEOs, and provide comparisons with non-resonant models. 
   These examples show that it is necessary to take into account the effect of the resonance for the computation of
   accurate proper elements. 
\end{abstract}

\section{Introduction}
It is well known that the gravitational $N$-body problem is non-integrable for $N \geq
3$, in the sense that there are not enough integrals of motion to find a complete set of
action-angle coordinates. In the context of our Solar System, the motion of an asteroid
can be treated as a perturbation of an integrable problem, i.e. the two-body problem
Sun-asteroid, and this permits to compute the so called \textit{proper elements}.  Proper
elements are quasi-integrals of motion, meaning that they are nearly constant in
time.
Another equivalent definition is that proper elements are actual
integrals of motion of an appropriately simplified model. These quantities represent the
average behavior of the orbit, thus they permit to describe the long term evolution of an
object using only few numerical parameters.

Proper elements have been computed first for main belt asteroids, and different techniques
have been used to this purpose \citep[see][for a historical
review]{knezevic-etal_2002, knezevic_2016}.
\citet{hirayama_1918, hirayama_1922} used the Lagrange linear theory of secular perturbations, and
determined proper elements of main-belt asteroids for the first time. He then used these
values to show that some asteroids are grouped in the space of proper orbital elements,
introducing the concept of asteroid families. 
\citet{brower_1951} was able to get more accurate proper elements by combining the linear
theory of secular perturbations and an improved theory of planetary motion.
Modern analytical methods \citep[see e.g.][]{milani-knezevic_1990, milani-knezevic_1992,
milani-knezevic_1994} are based on Hamiltonian perturbation theory. The perturbation is
expanded in Fourier series of cosines of combinations of the angular variables,
with amplitudes expressed as polynomials in eccentricity and sine of inclination.
However, this theory is efficient only when the eccentricity and the inclination are both
small. 
A secular analytical theory for high inclination orbits has been developed by
\citet{kozai_1962}, and proper elements computed with this technique have been provided by
\citet{kozai_1979}.
In order to overcome the limitation of small eccentricity and inclination assumption, one
can use semi-analytical methods.  \citet{williams_1969} introduced a method that avoids
the expansion in eccentricity and inclination, and provided a list of proper elements
computed with this technique in \citet{williams_1979, williams_1989}.  Later,
\citet{lemaitre-morbidelli_1994} extended this theory by using the Hamiltonian formalism. 
Finally, with the large performance improvements of electronic processors seen in the last
30 years, a completely numerical (or \emph{synthetic}) theory has been developed by
\citet{knezevic-milani_2000, knezevic-milani_2003}. 
This theory is based on a purely numerical integration, digital filtering for the removal
of short periodic oscillations, and Fourier analysis for the determination of proper
elements. On the other hand, the proper frequencies are determined by fitting the
evolution of the angular variables with a linear model.
Today, catalogs of main-belt asteroid proper elements are provided by the
AstDyS\footnote{\url{https://newton.spacedys.com/astdys/index.php?pc=0}} service, and by the
Asteroid Families Portal\footnote{\url{http://asteroids.matf.bg.ac.rs/fam/index.php}}
\citep{novakovic-radovic_2019}.
Data from these large catalogs have been successfully used to identify numerous asteroid 
families \citep[see e.g.][]{milani-etal_2014, nesvorny-etal_2015}, and to 
determine their ages \citep[see e.g.][]{vokrouhlicky-etal_2006, spoto-etal_2015}.

Differently from main belt asteroids, the orbits of NEOs can cross the orbit of one or
more planets, and for this reason the analytical and semi-analytical techniques mentioned
above can not be used in this context. Moreover, the Lyapunov time of the orbits of NEOs
is generally short, of the order of a few thousand years. Thus, also the synthetic theory
is not suitable for the computation of proper elements, because the integration of a
single orbit is not a good representative of the bulk of possible behaviors after a few
Lyapunov times.
The difficulty of the orbit crossing singularity has been addressed by
\citet{gronchi-milani_1998}. These authors considered the secular model defined by
\citet{kozai_1962}, where short term perturbations are removed by averaging over the fast
angles, and they introduced a technique to propagate the dynamics across a planet-crossing
configuration. 
Later, \citet{gronchi-milani_2001} used this method to compute proper elements of NEOs, that are
currently provided by NEODyS.\footnote{\url{https://newton.spacedys.com/neodys/}}
Despite their short time validity, NEOs proper elements can be used to search for
clusterings, and permit to identify a NEOs family that formed very recently.
A first search in the NEODyS catalog performed by \citet{schunova-etal_2012} did not 
produce any positive identification. However, about 27.000 NEOs are known today, which 
is about three times more than the number of NEOs known back in 2012. Additionally, 
once the Vera Rubin Observatory will start its operational lifetime, the number of known 
NEOs is expected to grow by a factor between 10 and 100 \citep{jones-etal_2016}, increasing
the chances for the first identification of an asteroid family among NEOs. 
Proper elements could also be used to identify secular resonances in the NEO region
\citep[][]{michel-froeschle_1997}, and provide a better understanding of the dynamical pathways
from the main belt to the NEO region.

NEOs proper elements defined by \citet{gronchi-milani_2001} are computed assuming that the
asteroid is not in a mean motion resonance with a planet.  However, mean motion resonances
with Jupiter play a fundamental role in delivering asteroids from the main belt to the NEO
region \citep[see e.g.][]{bottke-etal_2002, granvik-etal_2017}, and many NEOs are known to
be currently in resonance with a planet. In this dynamical setting, proper elements defined in
\citet{gronchi-milani_2001} are not appropriate, and a different model must be used. 
Here, we address the problem of the computation of proper elements for resonant NEOs.
Starting from the theory of \citet{gronchi-milani_2001}, we need to keep resonant
harmonics in the Hamiltonian by semi-averaging techniques, and the method must remain
valid for arbitrary eccentricities and inclinations. Semi-averaged theories have been
widely used in the past for describing the long-term orbital dynamics of resonant small
bodies \citep[see e.g][]{wisdom_1985, moons-morbidelli_1995, milani-baccili_1998,
saillenfest-etal_2016, saillenfest-etal_2017, sidorenko_2006, sidorenko_2018,
sidorenko_2020}.  Our aim is to combine such a method with the model of
\citet{gronchi-milani_2001} to compute proper elements of resonant NEOs. As the resonant
harmonics add one degree of freedom to the system, we must find a way to reduce the
problem to an integrable one, upon suitable transformations that are valid if secular
chaos is not too strong \citep[see e.g.][]{sidorenko-etal_2014}.  For this purpose, the traditional adiabatic approximation
\citep[see e.g.][]{neishtadt_1987, henrard_1993} may not be suitable for NEOs because of
the large orbital perturbations that they undergo. In such a case, the method of
\citet{knezevic-milani_2000} for synthetic proper elements can be applied to the
semi-averaged system, and proper frequencies can be computed by using a frequency analysis
\citep{laskar_1988, laskar_1990, laskar_2005}. 

The paper is structured as follows. 
In Sec.~\ref{s:semisecular} we introduce the semi-secular resonant model, and describe how to overcome the problem of 
the crossing singularity. 
In Sec.~\ref{s:propelNEO} we briefly describe the adiabatic invariant theory, that is
useful to better understand the secular evolution, but is shown to be quantitatively
inaccurate in the NEO region.
Then, we describe the use of frequency analysis for the determination
of proper frequencies and proper elements of NEOs.
In Sec.~\ref{s:examples} we provide some examples of known resonant NEOs, and we show the
differences in the results with respect to the non-resonant model by \citet{gronchi-milani_2001}.      
Finally, in Sec.~\ref{s:conclusions} we summarize the results.

\section{Averaging on resonant planet-crossing orbits}
\label{s:semisecular}

\subsection{The semi-secular evolution} 
Let us assume that the planets from Mercury to Neptune are placed on circular co-planar
orbits centered at the Sun. 
The same approximation has been used before by \citet{gronchi-milani_2001} for the
computation of proper elements of NEOs in the case of no mean-motion resonances, and
\citet{gronchi-michel_2001} verified that it is accurate enough for the description of the
secular motion of NEOs, provided that no close approaches with the planets occur.
We denote with $\k = \sqrt{\G m_0}$ the Gauss constant and set $\mu_j = m_j/m_0$, where $\G$ is the
universal gravitational constant, $m_0$ is the mass of the Sun and $m_j, j=1,\dots, 8$ are
the masses of the planets.
We introduce the Delaunay elements $(L,G,Z,\ell,g,z)$ of the asteroid as
\begin{equation}
\begin{split}
    L = \k\sqrt{a},              & \qquad \ell = M, \\
    G = \k\sqrt{a(1-e^2)},       & \qquad   g = \omega, \\
    Z = \k\sqrt{a(1-e^2)}\cos I, & \qquad   z = \Omega,
\end{split}
\end{equation}
where $a$ is the semi-major axis of the orbit of the asteroid, $e$ is the eccentricity, $I$ is the inclination,
$\Omega$ is the longitude of the ascending node, $\omega$ is the argument of the
pericenter, and $M$ is the mean anomaly.
In these coordinates, the Hamiltonian of the restricted problem is given by
\begin{equation}
    \H = \H_0 + \epsilon \H_1, \quad \epsilon=\mu_5,
\end{equation}
where $\H_0$ is the unperturbed Keplerian Hamiltonian of the asteroid, and $\H_1$ is the perturbation, i.e.
\begin{equation}
    \begin{split}
        \H_0 & = -\frac{\k^4}{2L^2}, \\
        \H_1 & = -\k^2\sum_{j=1}^8  \frac{\mu_j}{\mu_5}
        \bigg( 
        \frac{1}{|\mathbf{r}-\mathbf{r}_j|} - \frac{\mathbf{r} \cdot \mathbf{r}_j}{|\mathbf{r}_j|^3}
        \bigg).
    \end{split}
    \label{eq:hamiltonianTerms}
\end{equation}
In Eq.~\eqref{eq:hamiltonianTerms}, $\mathbf{r}$ and $\mathbf{r}_j, j=1,\dots,8$ are the
heliocentric positions of the asteroid and the planets, respectively.
Note that $\H$ directly depends on the time $t$, and because planets are on circular orbits we have 
$\ell_j = \n_j t + \ell_j(0)$, where $\ell_j, \n_j$ are the mean anomaly and the mean
motion of the $j$-th planet, respectively. To remove the direct time dependence we
over-extend the phase space by introducing a variable $L_j$ conjugated to $\ell_j$, 
so that we get an autonomous Hamiltonian. The Hamiltonian in the over-extended phase space is 
\begin{equation}
   \widetilde{\H} = \H_0 + \sum_{j=1}^8\n_j L_j + \epsilon \H_1,
\end{equation}
where $\H_1 = \H_1(L,L_1,...,L_8,G,Z,\ell,\ell_1,...,\ell_8,g,z)$. 
Let us assume that the asteroid is in a mean motion resonance with the 
$p$-th planet where the resonant angle is given by
\begin{equation}
    \sigma = h \lambda - h_p \lambda_p - (h-h_p)\varpi.
    \label{eq:resonantAngle}
\end{equation}
In Eq.~\eqref{eq:resonantAngle} the angles $\lambda=\ell+\omega+\Omega, \, \lambda_p=\ell_p+\omega_p+\Omega_p$ are the mean 
longitudes of the asteroid and the planet\footnote{Note that $\omega_p$ and
$\Omega_p$ are ill-defined because the planets move on circular and zero-inclination
orbits, hence we identify $\lambda_p$ with $\ell_p$.}, $\varpi = \omega+\Omega$ is the longitude of
the perihelion of the asteroid, and $h, \, h_p$ are co-prime
integers.\footnote{Eq.~\eqref{eq:resonantAngle} for $\sigma$ is used for the
purpose of definition, but since all resonant harmonics are kept in the Hamiltonian, this method 
describes all types of $h_p$:$h$ resonances at once (i.e. with a different combination of
$\Omega,\Omega_p, \varpi, \varpi_p$ fulfilling the D’Alembert rules)}
The integer number $|h-h_p|$ is usually called the resonance order. 
The Delaunay elements are first transformed into resonant semi-secular coordinates
\citep[see e.g.][]{saillenfest-etal_2016} by using the transformation
\begin{equation}
    \begin{pmatrix}
    \sigma \\
    \gamma \\
    u \\
    v \\
    \end{pmatrix} = 
    \begin{pmatrix}
    h & -h_p & h_p & h_p \\
    c & -c_p & c_p & c_p \\
    0 & 0    & 1   & 0   \\
    0 & 0    & 0   & 1   \\
    \end{pmatrix}
    \begin{pmatrix}
    \ell \\ \ell_p \\ g \\ z
    \end{pmatrix}, \qquad
    \begin{pmatrix}
    \Sigma \\
    \Gamma \\
    U \\
    V \\
    \end{pmatrix} = 
    \begin{pmatrix}
    -c_p & -c & 0 & 0 \\
    h_p & h & 0 & 0 \\
    0 & 1    & 1   & 0   \\
    0 & 1    & 0   & 1   \\
    \end{pmatrix}
    \begin{pmatrix}
    L \\ L_p \\ G \\ Z
    \end{pmatrix},
    \label{eq:canonicalTransformation}
\end{equation}
where $c,c_p$ are integers such that $c h_p - c_p h = 1$, that exist because $\text{gcd}(h,h_p)=1$.
Due to the resonance assumption, the angles $\ell_j, \ j=1,\dots,8$ and $\gamma$ evolve fast 
(frequency $\propto\epsilon^0$), the critical angle $\sigma$ evolves on a semi-secular
timescale (frequency $\propto \sqrt{\epsilon}$), and $(u,v)$ evolve on a secular
timescale (frequency $\propto \epsilon$).
Note that the quantities $\Gamma, L_j, j=1,\dots,8, \ j\ne p$ are first integrals in the
semi-secular coordinates, 
that have been introduced artificially by over-extending the phase space, and their value can be chosen arbitrarily. 
By choosing $\Gamma = 0$, the actions of Eq.~\eqref{eq:canonicalTransformation} become
\begin{equation}
    \Sigma = \frac{L}{h}, \qquad
    U = G - \frac{h_p}{h}L, \qquad
    V = Z - \frac{h_p}{h}L.
    \label{eq:semisecActions}
\end{equation}
The semi-secular Hamiltonian is obtained by averaging over the fast angles
\citep[see e.g.][]{milani-baccili_1998}, and it is given by
\begin{equation}
    \mathcal{K} = \mathcal{K}_0 + \epsilon(\mathcal{K}_{\text{sec}} + \mathcal{K}_{\text{res}}),
    \label{eq:semisecHamiltonian}
\end{equation}
where
\begin{equation}
    \begin{split}
        \mathcal{K}_0 & = -\frac{\k^4}{2(h\Sigma)^2} - \n_p h_p\Sigma, \\
        \mathcal{K}_{\text{sec}} & = 
        -\frac{\k^2}{\mu_5} \sum_{\substack{j=1 \\ j \ne p}}^8 \frac{\mu_j}{(2\pi)^2}\int_0^{2\pi}\int_0^{2\pi}\frac{1}{|\mathbf{r}-\mathbf{r}_j|}\text{d}\ell \text{d}\ell_j, \\
        \mathcal{K}_{\text{res}} & = -\frac{\k^2}{\mu_5} \frac{\mu_p}{2\pi}\int_0^{2\pi} \bigg( \frac{1}{|\mathbf{r}-\mathbf{r}_p|} - \frac{\mathbf{r}\cdot\mathbf{r}_p}{|\mathbf{r}_p|^3}\bigg)\text{d}\gamma.
    \end{split}
    \label{eq:semisecularHamiltonianTerms}
\end{equation}
The term $\mathcal{K}_{\text{sec}}$ contains the secular perturbations of all
the planets not involved in the resonance, while $\mathcal{K}_{\text{res}}$
contains all the resonant terms and the secular perturbation of the planet $p$.
The resonant normal form $\mathcal{K}$ does not depend on the angle 
$v$ in the semi-secular coordinates, 
because the problem is invariant with respect to rotations of the common orbital
plane of the planets \citep[see][]{kozai_1985, saillenfest-etal_2016}, hence the action $V$ is a
constant of motion. When the orbit of the asteroid does not cross the orbit of any 
planet, the Hamiltonian vector field defined by $\mathcal{K}$ can be
computed by exchanging the derivative and the integral sign. 
In case of orbit crossing this can not be done, because a singularity
appears in the integrals of the $\mathcal{K}_{\text{sec}}$ term \cite[see][]{gronchi-milani_1998}.

\subsection{Crossing singularity and numerical integration}
When the orbit of the asteroid and the orbit of a non-resonant planet intersect, the term
$\mathcal{K}_{\text{sec}}$ in Eq.~\eqref{eq:semisecularHamiltonianTerms} has a first-order
polar singularity, and solutions can be continued beyond the crossing.
A technique to extend solutions beyond the crossing singularity has been described in
\citet{gronchi-milani_1998, gronchi-tardioli_2013}. 
We summarize here the fundamental steps, leaving all the mathematical details in
Appendix~\ref{app:crossing}.

When the orbit of the asteroid and the orbit of the resonant planet intersect, the term
$\mathcal{K}_{\text{res}}$ has a collision singularity, occurring for a unique angle
$\sigma = \sigma_p$. In this case, the integral is divergent and solutions cannot be
continued beyond \cite[see e.g.][]{maro-gronchi_2018}.
This case never exactly occurs in practice and we will not mention it further in this
paper.

\subsubsection{Extraction of the singularity}
Let us suppose that there is only one non-resonant planet, and denote with $\mathbf{r}'$ its
heliocentric position. Then
\begin{equation}
   \mathcal{K}_{\text{sec}} \propto 
   \int_{0}^{2\pi}\int_{0}^{2\pi}\frac{1}{d} \,
   \text{d}\ell\text{d}\ell',
   \label{eq:simpleKsec}
\end{equation}
where $d = |\mathbf{r}-\mathbf{r}'|$. 
Let $y \in \{ \Sigma, U,V,\sigma,u,v \}$ be one of the coordinates, then the Hamiltonian vector field
is determined by the derivatives 
\begin{equation}
   \frac{\partial}{\partial y}\int_{0}^{2\pi}\int_{0}^{2\pi}\frac{1}{d} \,
   \text{d}\ell\text{d}\ell'.
   \label{eq:simpleDer}
\end{equation}
When the orbit of the asteroid and the orbit of the planet do not intersect, it is
possible to compute the derivative in Eq.~\eqref{eq:simpleDer} by exchanging the derivative
and the integral sign.
On the other hand, in the case of orbit crossing this can not be done, because the double
integral has a polar singularity.
In a neighborhood of the crossing configuration, a function $\delta$ that
approximates the distance $d$ is
defined by using the minimum orbit intersection distance \citep[MOID;][]{gronchi_2005} between 
the orbit of the planet and that of the asteroid.
The integral in Eq.~\eqref{eq:simpleDer} is therefore decomposed as
\begin{equation}
   \frac{\partial}{\partial y}\int_{0}^{2\pi}\int_{0}^{2\pi}\frac{1}{d} \,
   \text{d}\ell\text{d}\ell' = 
   \frac{\partial}{\partial y}\int_{0}^{2\pi}\int_{0}^{2\pi}\frac{1}{\delta} \,
   \text{d}\ell\text{d}\ell'  + 
   \frac{\partial}{\partial y}\int_{0}^{2\pi}\int_{0}^{2\pi}\bigg(\frac{1}{d} -
   \frac{1}{\delta}\bigg) \,
   \text{d}\ell\text{d}\ell'.
   \label{eq:simpleKantorovich}
\end{equation}
It turns out that the second term in the right hand side of
Eq.~\eqref{eq:simpleKantorovich} can be computed numerically by exchanging the derivative
and the integral signs.
The first term in the right hand side of Eq.~\eqref{eq:simpleKantorovich} contains the
principal part of the singularity of the term in the left hand side, with the advantage that it can be computed with
an analytical formula (see Appendix~\ref{app:crossing}). Moreover, the discontinuity is given
only by the discontinuity in the derivatives of the MOID. This analytical formulation
permits to compute the difference between the Hamiltonian vector field on the two sides
of the orbit crossing singularity (see also Eq.~\ref{eq:derJump}), enabling us to continue a
solution beyond these singular configurations.

Note that a double crossing can occur. In this case, we can define two functions
$\delta_1, \delta_2$ approximating the distance $d$ in a neighbourhood of the two
crossing points, and we can extract the singularities using the decomposition
\begin{equation}
   \frac{1}{d} = \frac{1}{\delta_1} + \frac{1}{\delta_2} +  \bigg( \frac{1}{d} -
   \frac{1}{\delta_1} - \frac{1}{\delta_2} \bigg) .
\end{equation}

\subsubsection{Numerical integration scheme}
\label{ss:initCond}
The numerical implementation of the continuation of solutions beyond an orbit crossing follows that of
\citet{gronchi-milani_2001}. We use an implicit Runge-Kutta-Gauss scheme
\citep[see e.g.][]{hairer-etal_2002} for the integration of the equations of motions 
generated by the semi-secular Hamiltonian $\mathcal{K}$. Jacobi iterations are used to
solve the fixed-point equation needed to compute the coefficients of the Runge-Kutta-Gauss
method, and the first guess is computed by using a polynomial extrapolation from the previous integration point.

The MOID with each planet is computed at every integration step, and these values are used to check whether a
crossing with a non-resonant planet has occurred or not. 
If a planet crossing is detected, an iterative method is used to make the integration step
arrive exactly at the crossing configuration.  When this singular configuration is reached
within the required precision, the integration needs to restart from this point for the
next step. 
\begin{figure}[!ht]
   \centering
   \includegraphics[width=0.7\textwidth]{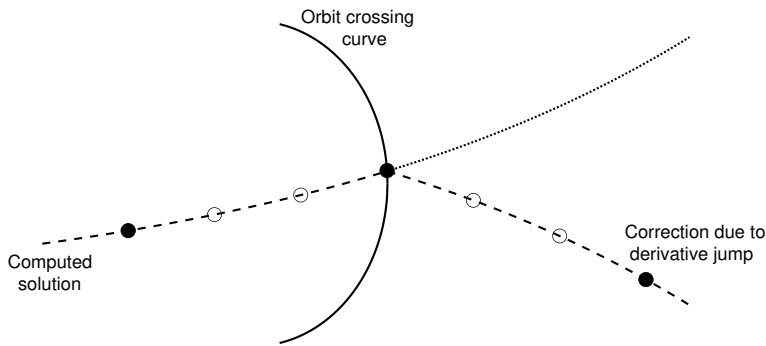}
   \caption{Graphical description of the Runge-Kutta-Gauss method used for the computation
      of solutions beyond the orbit crossing. 
      In this example, a Runge-Kutta-Gauss of order 6 is presented. 
      The dashed curve is the solution of the system, and black filled circles are the 
      points at which the solution is computed. Empty circles are the two additional 
      intermediate points needed for the Runge-Kutta-Gauss method to compute the
      solution, and they are the only points at which the force is computed.
      The dotted curve represents the solution of the system after the orbit crossing,
      obtained by correcting the vector field with the jump in the derivatives.
   }
   \label{fig:firstGuessCorrection}
\end{figure}
In this case, the first guess for the Jacobi iteration computed with polynomial
extrapolation would be wrong, since there is a jump in the vector field at the crossing
curve. A good first guess is therefore computed by correcting the polynomial extrapolation 
using the analytical formula of the jump of Eq.~\eqref{eq:derJump}.
Figure~\ref{fig:firstGuessCorrection} describes this integration scheme.

The semi-secular Hamiltonian Eq.~\eqref{eq:semisecHamiltonian} 
is a function of the semi-secular coordinates defined in
Eq.~\eqref{eq:canonicalTransformation}, hence osculating orbital elements should not be used
as initial conditions.
In order to compute the appropriate semi-secular coordinates for a given small body, we
integrate its orbit in a full $N$-body problem for 1000 yr, including all the planets from
Mercury to Neptune in the model. Short periodic oscillations are then removed through a
digital filter \citep{carpino-etal_1987}. 
These steps are performed with the \textsc{orbit9} integrator, included in the 
\textsc{OrbFit}\footnote{\url{http://adams.dm.unipi.it/orbfit/}} package.
Filtered elements at the initial time are then used as initial conditions for the
semi-secular Hamiltonian.
For the \textsc{orbit9} integrations performed in this paper, we used initial conditions
for the planets at time 59000 MJD, taken from the JPL
Horizons\footnote{\url{https://ssd.jpl.nasa.gov/?horizons}} ephemeris system.  Orbital
elements of NEOs at time 59000 MJD were taken from the NEODyS catalog, and the nominal
orbits were used as initial condition. To propagate the orbits, we used the Everhart integration method
\citep{everhart_1985}.

\section{Proper elements of resonant NEOs}
\label{s:propelNEO}
\subsection{Adiabatic approximation}
\label{s:adiabatic}
The variables of the Hamiltonian system defined by Eq.~\eqref{eq:semisecHamiltonian} evolve on two
different timescales. The couple $(\Sigma, \sigma)$ evolves over a semi-secular timescale,
while $(U,u)$ evolves over a secular timescale, and under suitable conditions this separation can
be used to further simplify the problem. Denote with $\nu_\sigma$ (that is
$\propto\sqrt{\epsilon}$) and $\nu_u$ (that is $\propto \epsilon$) the
frequencies associated to these two couples of variables, and assume that
\begin{equation}
   \xi = \frac{\nu_u}{\nu_\sigma} \ll 1.
   \label{eq:adiabaticApproximation}
\end{equation}
If condition \eqref{eq:adiabaticApproximation} holds, the adiabatic invariant theory \citep[see e.g.][]{henrard_1993} can be used 
to transform $(\Sigma, \sigma)$ into a new pair of action-angle coordinates $(J,\theta)$. In
this manner, we reduce the problem to a system with one-degree of freedom. 
The momentum $J$ is also called the \emph{adiabatic invariant} and, although it is not exactly
conserved in the semi-secular system, its variations can be discarded for a sufficiently
small $\xi$. 
This approach has been used to understand the qualitative dynamics of objects in the 
strongest mean-motion resonances with Jupiter \citep[see e.g.][]{wisdom_1985,
henrard-lemaitre_1987, sidorenko_2006}, and to study co-orbital
motions \citep[see e.g.][]{sidorenko-etal_2014, sidorenko_2020}.
The adiabatic invariant is defined as the area enclosed (or stretched) 
by an orbit $\big(\Sigma(t), \sigma(t)\big)$ (also called \emph{guiding trajectory}) computed
keeping the variables $(U,u)$ fixed, and in the libration case it is given by
\begin{equation}
    2\pi J = \frac{1}{2}\oint\big( \Sigma \text{d}\sigma - \sigma \text{d}\Sigma \big) 
    = \frac{1}{2} \int_0^{T_\sigma} \big( \dot{\sigma} \Sigma - \dot{\Sigma}\sigma \big) \text{d}t,
    \label{eq:2piJ}
\end{equation}
where $T_\sigma$ is the period of the guiding trajectory. 
If $(\Sigma_0, \sigma_0)$ is any point of the orbit, then the secular Hamiltonian describing 
the evolution of the couple $(U,u)$ is given by
\begin{equation}
   \mathcal{F}(J,U,V,\theta, u) = \mathcal{K}(\Sigma_0, U, V, \sigma_0, u) +
   \mathcal{O}(\epsilon^{3/2}).
    \label{eq:secularHamiltonian}
\end{equation}
If we neglect the remainder, the secular Hamiltonian of Eq.~\eqref{eq:secularHamiltonian} has one degree of freedom,
and the solutions follow its level curves in the $(U,u)$-plane.
This secular model has been successfully used by \citet{saillenfest-etal_2016,
saillenfest-etal_2017, saillenfest-lari_2017} to study the long-term evolution of distant
resonant trans-Neptunian objects (TNOs), for which the ratio $\xi$ is of the order of $10^{-4}$.

\begin{figure}[!ht]
   \centering
   \includegraphics[width=\textwidth]{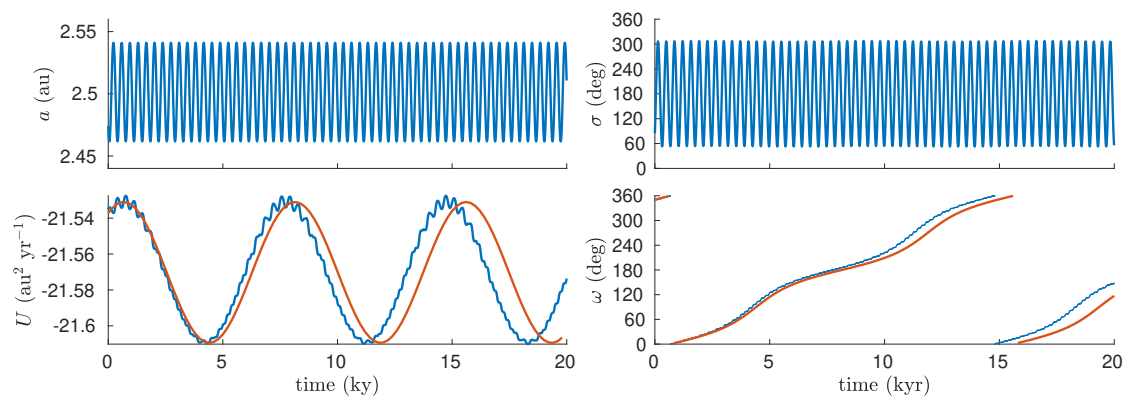}
   \caption{Comparison of the semi-secular dynamics (blue curve) and secular dynamics assuming the
adiabatic invariance (red curve). The asteroid taken as example here is (887) Alinda.}
   \label{fig:887_a_sigma}
\end{figure}

However, NEOs evolve on a much shorter timescale than TNOs, and the condition in  
Eq.~\eqref{eq:adiabaticApproximation} may not be well satisfied in general.
To show that the secular model of Eq.~\eqref{eq:secularHamiltonian} is not suitable to get
accurate proper elements in the context of NEOs, we consider the example of asteroid (887)
Alinda, that is currently in a 3:1 mean motion resonance with Jupiter. 
We numerically integrated its dynamics using both the semi-secular Hamiltonian of
Eq.~\eqref{eq:semisecHamiltonian}, and the secular one of
Eq.~\eqref{eq:secularHamiltonian} assuming the adiabatic invariance hypothesis. 
Figure~\ref{fig:887_a_sigma} shows the semi-secular evolution of the semi-major axis $a$ and of the
critical argument $\sigma$ for 20 ky, and the comparison between the evolution computed
with the semi-secular and the secular models.
(887) Alinda stays in the resonance for the whole
integration timespan, and the period of the guiding trajectory is
approximately $T_\sigma \approx 360$ yr.
As expected, the semi-secular evolution has small oscillations
with a period equal to $T_\sigma$, while in the secular evolution only the main
long-term oscillation is kept. 
The secular period here is approximately $T_u \approx 14200$ yr, resulting in a ratio of
$\xi \approx
0.025$, almost three orders of magnitude larger than in the case of distant TNOs. This already
suggests that the adiabatic approximation might not be appropriate for an accurate
computation of proper elements.
More importantly, from Fig.~\ref{fig:887_a_sigma} we can notice that, although the secular
model is able to reproduce correctly the amplitude of oscillation of $U$ and explain the
dynamics of Alinda in a qualitative way, the frequency is not the same as the one obtained
in the semi-secular evolution.  This shows that the use of the adiabatic approximation for
the computation of the secular Hamiltonian could lead to a poor determination of the
proper elements, in particular of the proper frequencies for the case of (887) Alinda.
Thus, we must find another way to extract the proper elements from the resonant normal
form $\mathcal{K}$.

\subsection{Frequency analysis}
\label{s:properel}
Proper frequencies are an intrinsic property of the dynamical system 
in its integrable approximation. In other words, even though the semi-secular system has two 
degrees of freedom, it already contains the proper frequencies of the asteroid, and one must 
just find a way to properly extract them. In the absence of a well defined adiabatic invariant,
the fundamental frequencies of the system can be computed numerically by performing a frequency 
analysis on the semi-secular time series.

To this end, the dynamics of a resonant NEO is propagated forward in time for 200 kyr using the semi-secular
model of Sec.~\ref{s:semisecular}, and the time series of the resonant elements $(\Sigma, U, V, \sigma, u, v)$ 
are converted to the time series of the Keplerian elements $e, \cos I, \omega, \Omega$. 
Then, we apply the frequency analysis method by J. Laskar \citep[see][]{laskar_1988,
laskar_1990, laskar_2005}
to the functions
\begin{equation}
   \eta = e \exp(i\omega), \qquad \zeta = \sin\frac{I}{2}\exp(i\Omega),
   \label{eq:functionNAFF}
\end{equation}
and determine their frequency decomposition. We perform the frequency analysis using the
TRIP\footnote{\url{https://www.imcce.fr/Equipes/ASD/trip/trip.php}} software developed by
\citet{gastineau-laskar_2011}.

The functions $\eta$ and $\zeta$ are expressed as quasi-periodic series, in which the
frequency of each term is an integer combination of the proper frequencies $\nu_\sigma,
\nu_u$, and $\nu_v$ \citep[see e.g.][]{laskar-etal_1992}. Therefore, we need to identify
the proper frequencies from their integer combinations. Since the semi-secular Hamiltonian
is invariant by rotation (and thus has only two degrees of freedom; see
Sec.~\ref{s:semisecular}), the decomposition of $\eta$ only features $\nu_\sigma$
and $\nu_u$. The frequency $\nu_u$ is usually that of the largest-amplitude term of $\eta$
(as it is the case for Alinda, see Fig.~\ref{fig:887_a_sigma}). However, in order to avoid
any ambiguity, the frequency $\nu_\sigma$ can be determined by a preliminary frequency
analysis of the variables $(\Sigma,\sigma)$. Once we have $\nu_\sigma$ and $\nu_u$, the
identification of $\nu_v$ from the analysis of $\zeta$ is straightforward. The proper
frequencies $g-s$ of the argument of the pericenter $\omega$, and $s$ of the longitude of
the node $\Omega$, correspond to $\nu_u$ and $\nu_v$, respectively.  Note that if the
quasi-periodic decomposition of $\eta$ contains a constant term, then $\omega$ librates,
otherwise it circulates. 
As an example, Table~\ref{tab:887_freq} gives the terms of the quasi-periodic
decomposition of $\zeta$ for (887) Alinda.
\begin{table}[htbp!]
   \centering
   \renewcommand{\arraystretch}{1.2}
   \caption{Frequencies, amplitudes, and phases of the quasi-periodic decomposition of the
   function $\zeta$ for asteroid (887) Alinda. The table shows only the terms with an
      amplitude larger than $10^{-4}$.}
   \begin{tabular}{|r|r|r|r|}
      \hline
      Frequency identification & Frequency (arcsec yr$^{-1}$) & Amplitude & Phase (deg)
      \\
      \hline
      \hline
$                           \nu_v $ &   $-$80.92578  & 6.43469$\cdot10^{-2}$  & 119.027 \\
$                 2\nu_u +  \nu_v $ &   103.00656  & 1.84703$\cdot10^{-2}$  &  82.034 \\
$  \nu_\sigma  +  2\nu_u +  \nu_v $ &  3723.24145  & 6.61366$\cdot10^{-4}$  &  25.051 \\ 
$ -\nu_\sigma            +  \nu_v $ & $-$3701.11622  & 6.20626$\cdot10^{-4}$  & 173.959 \\
$  \nu_\sigma            +  \nu_v $ &  3539.21386  & 6.18166$\cdot10^{-4}$  &  65.097 \\
$ -\nu_\sigma  +  2\nu_u +  \nu_v $ & $-$3517.05867  & 6.00843$\cdot10^{-4}$  & 135.966 \\
$  2\nu_\sigma           +  \nu_v $ &  7159.40976  & 2.97608$\cdot10^{-4}$  &  15.804 \\
$ -2\nu_\sigma           +  \nu_v $ & $-$7321.18962  & 2.59903$\cdot10^{-4}$  &  50.200 \\
$  2\nu_\sigma +  2\nu_u +  \nu_v $ &  7343.45527  & 1.88732$\cdot10^{-4}$  & 149.424 \\
$ -2\nu_\sigma +  2\nu_u +  \nu_v $ & $-$7137.08519  & 1.59706$\cdot10^{-4}$  & 176.425 \\
      \hline
   \end{tabular}
   \label{tab:887_freq}
\end{table}

In order to complete the set of proper elements, we now need to compute the central value
and the amplitude of secular oscillations of eccentricity and inclination. To this
purpose, we remove all the semi-fast terms from the quasi-periodic decomposition (i.e. those
featuring the frequency $\nu_\sigma$). Then, we obtain the bounds of the secular variation of
eccentricity and inclination by summing up or subtracting the amplitudes of the terms
in the quasi-periodic series. The last term of the series (i.e. the term with smallest
amplitude) sets the accuracy of these estimates. It depends on the capability of the
frequency analysis algorithm to express the signal as a quasi-periodic series: proper
elements are very precise for nearly-integrable semi-secular trajectories, but poorly
determined if secular chaos is strong. As explained in Sec.~\ref{s:examples}, secular chaos
is an intrinsic limitation for the computation of proper elements.

\section{Examples}
\label{s:examples}
\begin{table}
   \renewcommand{\arraystretch}{1.3}
\setlength{\tabcolsep}{2pt}
  \begin{threeparttable}
   \centering
   \caption{Proper elements of some selected resonant NEOs.}
   {\scriptsize
   \begin{tabular}{|c|c|c|c|c|c|c|c|c|c|c|}
      \hline
      Des. & $h_p$:$h$ & Res. & Cross & $(e_{\min}, e_{\max})$ &
      $(I_{\min}, I_{\max})$ & $g-s$ & $s$ & $lf$ & $(\omega_{\min}, \omega_{\max})$ & Case \\
       &  & pla. & pla. &  & &  & & & & \\
      \hline
      \hline
(887)     & 3:1 & J & M     & (0.5524, 0.5641) & (5.29,  9.49) &   91.97 &  $-$80.93 & / & / & A2\\ 
(2608)    & 3:1 & J & M     & (0.5282, 0.5769) & (10.94, 19.18) &  216.30 & $-$150.11 & / & / &A2\\ 
(8201)    & 3:1 & J & E/M   & (0.7140, 0.7296) & (5.39, 13.47) &   81.05 &  $-$72.42 & / & / &A2\\ 
(19356)   & 3:1 & J & M     & (0.5637, 0.5656) & (1.94,  3.74) &   45.18 &  $-$56.06 & / & / &A2\\ 
(153311)  & 3:1 & J & E/M   & (0.3444, 0.7098) & (17.64, 44.37) &   17.75 &  $-$30.80 & / & / &A2\\ 
(6178)    & 5:2 & J & M     & (0.5747, 0.5812) & (3.67,  7.10) &   70.67 &  $-$71.27 & / & / &A2\\ 
(26760)   & 5:2 & J & M     & (0.5589, 0.5830) & (6.89, 13.42) &  135.14 & $-$115.11 & / & / &A2\\ 
(14827)   & 5:2 & J & E/M   & (0.6712, 0.6728) & (2.32,  4.33) &  385.83 & $-$315.24 & / & / &A2\\ 
(152667)  & 5:2 & J & E/M   & (0.6986, 0.7051) & (3.70,  8.51) &  231.96 & $-$185.15 & / & / &A2\\ 
(481482)  & 5:2 & J & V/E/M & (0.7550, 0.8135) & (12.84, 30.09) &  295.95 & $-$259.61 & / & / &A2\\ 
(34613)   & 4:1 & J & M     & (0.3783, 0.4047) & (6.04,  7.95) &   86.62 &  $-$46.97 & / & / &A2\\ 
(361518)  & 4:1 & J & E/M   & (0.5972, 0.6021) & (4.15,  6.87) &   37.18 &  $-$45.67 & / & / &A2\\ 
(369983)  & 4:1 & J & M     & (0.3854, 0.3856) & (1.02,  1.33) &   94.55 &  $-$51.68 & / & / &A2\\ 
(407653)  & 4:1 & J & M     & (0.4846, 0.5036) & (8.26, 12.18) &   69.67 &  $-$51.10 & / & / &A2\\ 
(408752)  & 4:1 & J & V/E/M & (0.7768, 0.7832) & (3.93,  9.90) &   91.14 &  $-$80.05 & / & / &A2\\ 
(303174)  & 7:2 & J & M     & (0.3212, 0.4503) & (21.28, 28.34) &   57.61 &  $-$43.20 & / & / &A2\\ 
(329395)  & 7:2 & J & E/M   & (0.3900, 0.7180) & (25.82, 47.11) &   46.84 &  $-$45.14 & / & / &A2\\ 
(452639)  & 7:2 & J & V/E/M & (0.8698, 0.8755) & (3.46, 12.16) &  157.70 & $-$135.86 & / & / &A2\\ 
(488494)  & 7:2 & J & M     & (0.4474, 0.4669) & (7.87, 11.31) &   92.21 &  $-$62.10 & / & / &A2\\ 
(501878)  & 7:2 & J & V/E/M & (0.7543, 0.7859) & (7.88, 21.08) &  124.45 & $-$105.99 & / & / &A2\\ 
(4503)    & 8:3 & J & M     & (0.5210, 0.5238) & (2.51,  4.41) &  218.75 & $-$159.75 & / & / &A2\\ 
(9172)    & 8:3 & J & M     & (0.5290, 0.5558) & (7.67, 13.87) &  144.70 & $-$117.00 & / & / &A2\\ 
(152575)  & 8:3 & J & M     & (0.5258, 0.5487) & (7.04, 12.72) &  205.91 & $-$151.58 & / & / &A2\\ 
(363076)  & 8:3 & J & M     & (0.4884, 0.5174) & (8.14, 13.89) &  206.32 & $-$145.97 & / & / &A2\\ 
(405562)  & 8:3 & J & M     & (0.5393, 0.5430) & (2.96,  5.25) &  133.69 & $-$111.53 & / & / &A2\\ 
(416804)  & 8:3 & J & M     & (0.5670, 0.5793) & (4.99,  9.66) &  208.73 & $-$161.99 & / & / &A2\\ 
(5370)    & 2:1 & J & M/J   & (0.6280, 0.6805) & (10.49, 22.20) &   51.15 &  $-$38.04 & / & / &A2\\ 
(26166)   & 2:1 & J & M     & (0.5656, 0.6594) & (12.22, 27.02) & / &  $-$19.57 & 30.83 & (55.85, 124.15) &A2\\ 
(523592)  & 2:1 & J & /     & (0.3891, 0.5859) & (18.11, 33.25) &   73.95 &  $-$75.14 & / & / &A2\\ 
1999SE10  & 2:1 & J & M     & (0.6029, 0.6098) & (3.11,  7.31) &   32.01 &  $-$30.74 & / & /  &A2\\ 
2005YC    & 2:1 & J & M     & (0.4848, 0.6379) & (18.84, 33.57) & / &  $-$30.74 & 52.92 & (53.12, 126.88) &A2\\ 
2008KD6   & 2:3 & E & E/M   & (0.3799, 0.4710) & (21.98, 27.83) &   24.74 &  $-$17.65 & / & /  &A2\\ 
(358744)  & 2:5 & E & E/M   & (0.5121, 0.5139) & (3.08,  4.18) &   72.10 &  $-$48.92 & / & / &A2\\ 
2008EG9   & 3:5 & E & E/M   & (0.3855, 0.3861) & (2.45,  2.80) &   77.61 &  $-$53.16 & / & / &A2 \\ 
(10302)   & 3:7 & V & /     & (0.1369, 0.1403) & (4.01,  4.38) &   50.53 &  $-$28.11 & / & / &A2\\ 
2014HU46  & 1:3 & V & E/M   & (0.4188, 0.4191) & (1.52,  1.87) &   78.72 &  $-$70.04 & / & / &A2 \\ 
2012DF4   & 1:3 & V & V/E/M & (0.5382, 0.5395) & (4.61,  5.26) &   48.01 &  $-$38.41 & / & / &A2 \\ 
2015XU351 & 2:7 & V & V/E/M & (0.5761, 0.6084) & (11.39, 17.86) &   51.10 &  $-$39.67 & / & / &A2 \\ 
2016AH9   & 2:7 & V & E/M   & (0.5552, 0.5559) & (2.97,  3.51) &   62.66 &  $-$49.85 & / & /  &A2\\ 
(5381)    & 2:3 & V & V/E   & (0.1268, 0.6600) & (33.42, 50.79) & / &   $-$9.25 & 14.40 & (37.28, 142.73) &A2\\ 
(138911)  & 6:5 & M & /     & (0.0816, 0.0818) & (1.64,  1.68) &   33.07 &  $-$19.30 & / & / &A2\\ 
(159560)  & 4:7 & M & /     & (0.2650, 0.4380) & (25.50, 32.70) &  45.411  &  $-$36.32 & / & / &A1\\
(163412)  & 5:7 & M & M     & (0.3166, 0.5562) & (27.70, 39.13) &   35.02 &  $-$30.00 & / & / &B1\\ 
(208565)  & 4:3 & M & V/E/M & (0.4316, 0.8229) & (29.01, 56.56) &   17.98 &  $-$18.25 & / & / &A2\\ 
(211871)  & 3:4 & M & E/M   & (0.5107, 0.5185) & (5.64,  8.18) &   65.01 &  $-$46.95 & / & / &B1\\ 
(309728)  & 5:7 & M & M     & (0.3406, 0.4121) & (18.31, 23.07) &   46.98 &  $-$31.87 & / & / &B1\\ 
(10636)   & 8:11& M & M    & (0.4676, 0.5148) & (13.24, 19.25) &   58.49 &  $-$40.87 & / & / &B1\\ 
     \hline
   \end{tabular}
}
   \label{tab:NEOlist}
\begin{tablenotes}
      \scriptsize
   \item \textbf{Note}. Planets in the third and fourth column are indicated only with
      their initial letter (V = Venus, E = Earth, M = Mars, J = Jupiter). 
      The inclinations $I_{\min}, I_{\max}$ and the arguments 
      $\omega_{\min}, \omega_{\max}$ are in degrees, while the frequencies
      $g-s, s, lf$ are in arcsec yr$^{-1}$, where $lf$ is the oscillation frequency of
      $\omega$ in the case of Kozai resonance.
    \end{tablenotes}
  \end{threeparttable}
\end{table}

Table~\ref{tab:NEOlist} lists the NEOs that we took into account in this work. First, we
performed pure $N$-body simulations to check that both the critical angle 
$\sigma$ and the semi-major axis $a$ are currently oscillating for the indicated
resonance. 
Then, we computed proper elements using the method described in Sec.~\ref{s:properel}. 
We classify the examples according to the following cases:
\begin{itemize}
   \item[\textbf{A1}] No separatrix crossings, unnoticeable effect of the resonance;
   \item[\textbf{A2}] No separatrix crossings, strong effect of the resonance;
   \item[\textbf{B1}] Separatrix crossings, regular secular motion;
   \item[\textbf{B2}] Separatrix crossings, chaotic secular motion.
\end{itemize}
Separatrix crossings occur when the asteroid is pushed outside of the
resonance. The resonant angle $\sigma$ therefore switches from libration to circulation.
Separatrix crossings may greatly alter the secular evolution and produce long-term
chaos, even though the dynamics may be perfectly adiabatic between each crossing
event. This phenomenon is further described below.

\subsection{Case A1}
\begin{figure}[!ht]
   \centering
   \includegraphics[width=0.48\textwidth]{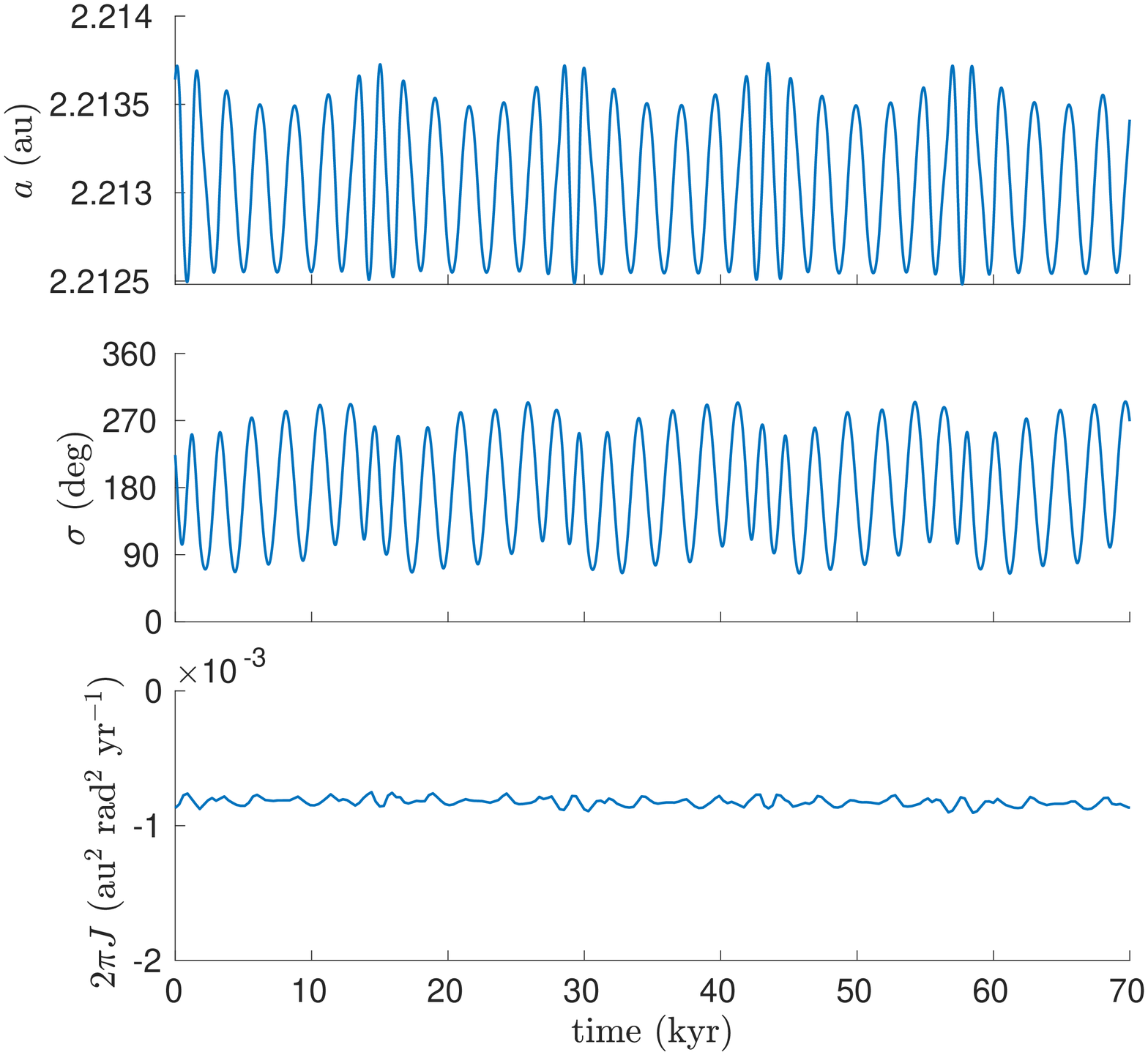}
   \includegraphics[width=0.48\textwidth]{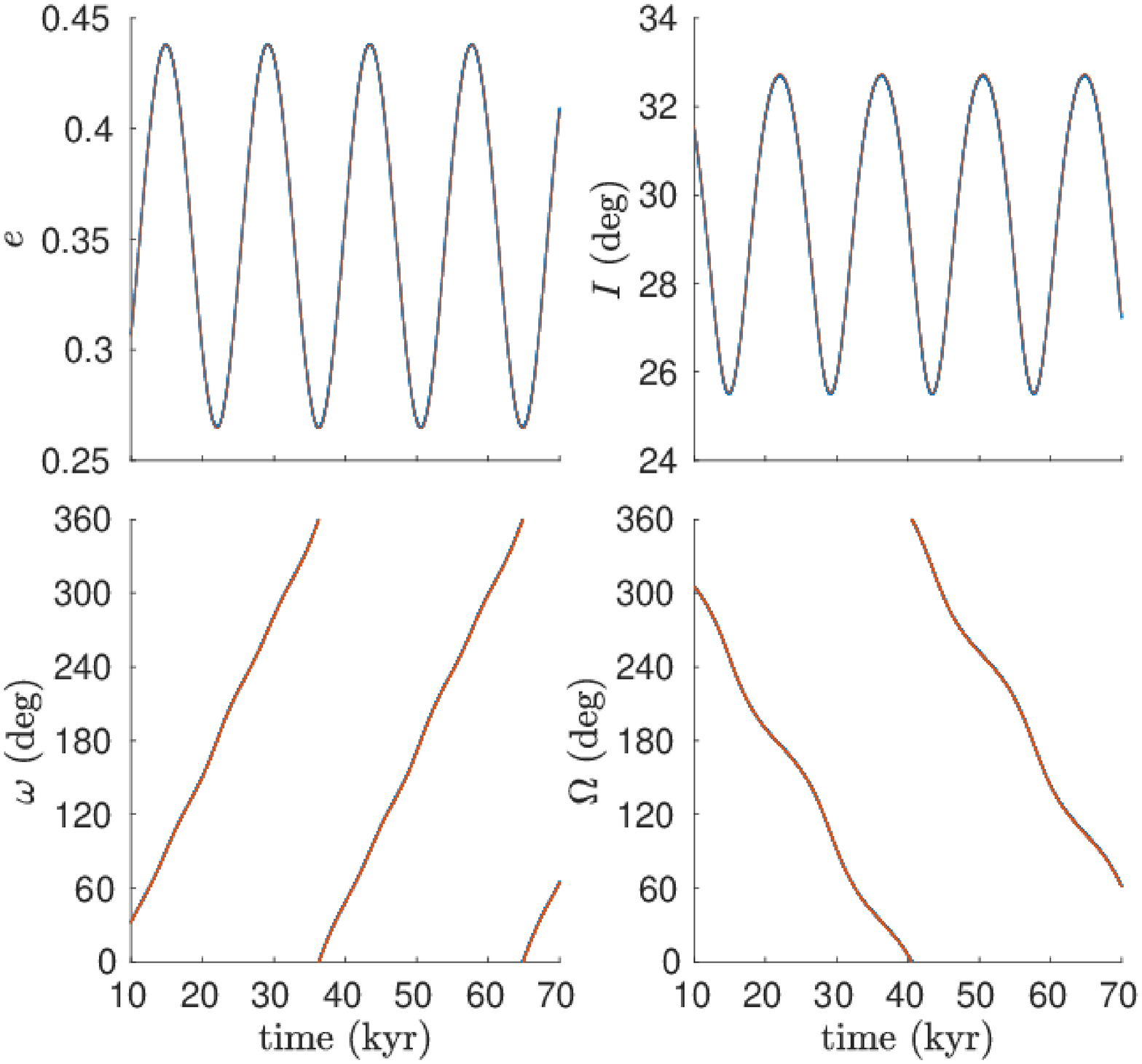}
   \caption{Long-term dynamics of asteroid (159560) 2001TO103. 
   The blue curve is the evolution obtained by the semi-secular model. 
   In the four rightmost panels, all terms with period smaller than 10 kyr 
   have been digitally filtered, and the red curve is the evolution obtained 
   with the secular non-resonant model of \citet{gronchi-milani_2001}. 
   The blue and red curves almost overlap and can hardly be distinguished one from the other.}
   \label{fig:159560_adiabatic}
\end{figure}

We first discuss the example of (159560) 2001 TO103, an asteroid in 4:7 mean motion
resonance with Mars that does not cross any planet. We propagated the semi-secular
dynamics for 200 ky, and we also monitored the evolution of the adiabatic invariant.  The
value $2\pi J$ is computed every 300 yr by using Eq.~\eqref{eq:2piJ}, where the variables
$(\Sigma, U, V, \sigma, u, v)$ identifying the guiding trajectory are taken from the
integration of the semi-secular dynamics.  
The results are shown in Fig.~\ref{fig:159560_adiabatic}.
The asteroid stays in the resonance for the whole integration timespan, and
the adiabatic invariant $2\pi J$ remains fairly constant (see
Fig.~\ref{fig:159560_adiabatic}, bottom left panel), experiencing a maximum relative
variation of only 10\% with respect to the initial value.
By looking at the orbital evolution, we see that the period of the guiding trajectories
lies somewhere between $T_\sigma = 1500$ and $2500$ yr, while the period of
circulation of
$\omega$ is $T_u \approx 30000$ yr, corresponding to a ratio between $\xi \approx 0.05$
and $\xi \approx 0.083$. 

The proper elements reported by NEODyS and obtained without taking the resonance into
account are $(e_{\min}, e_{\max}) = (0.2649, 0.4385)$, $(I_{\min}, I_{\max}) =
(25.522^\circ, 32.749^\circ)$, and the proper frequencies are $g-s =45.419 $ arcsec
yr$^{-1}$, $s =-36.367$ arcsec yr$^{-1}$.  Figure~\ref{fig:159560_adiabatic} shows the
comparisons of the evolution of $e, I, \omega$, and $\Omega$ obtained when taking the
resonance into account in the model or not.
The two dynamics are essentially the same, meaning that the effects of the resonance are
not noticeable on the long-term dynamics.  Indeed, the resonant proper elements that we
obtained in Table~\ref{tab:NEOlist} are very close to those reported by NEODyS; this
confirms that the mean-motion resonance does not have a strong influence on this object.
This also shows that our method, while taking the resonance into account,
is as precise as that of \citet{gronchi-milani_2001} for proper elements computation.

\subsection{Case A2}
\begin{figure}[!ht]
   \centering
   \includegraphics[width=0.48\textwidth]{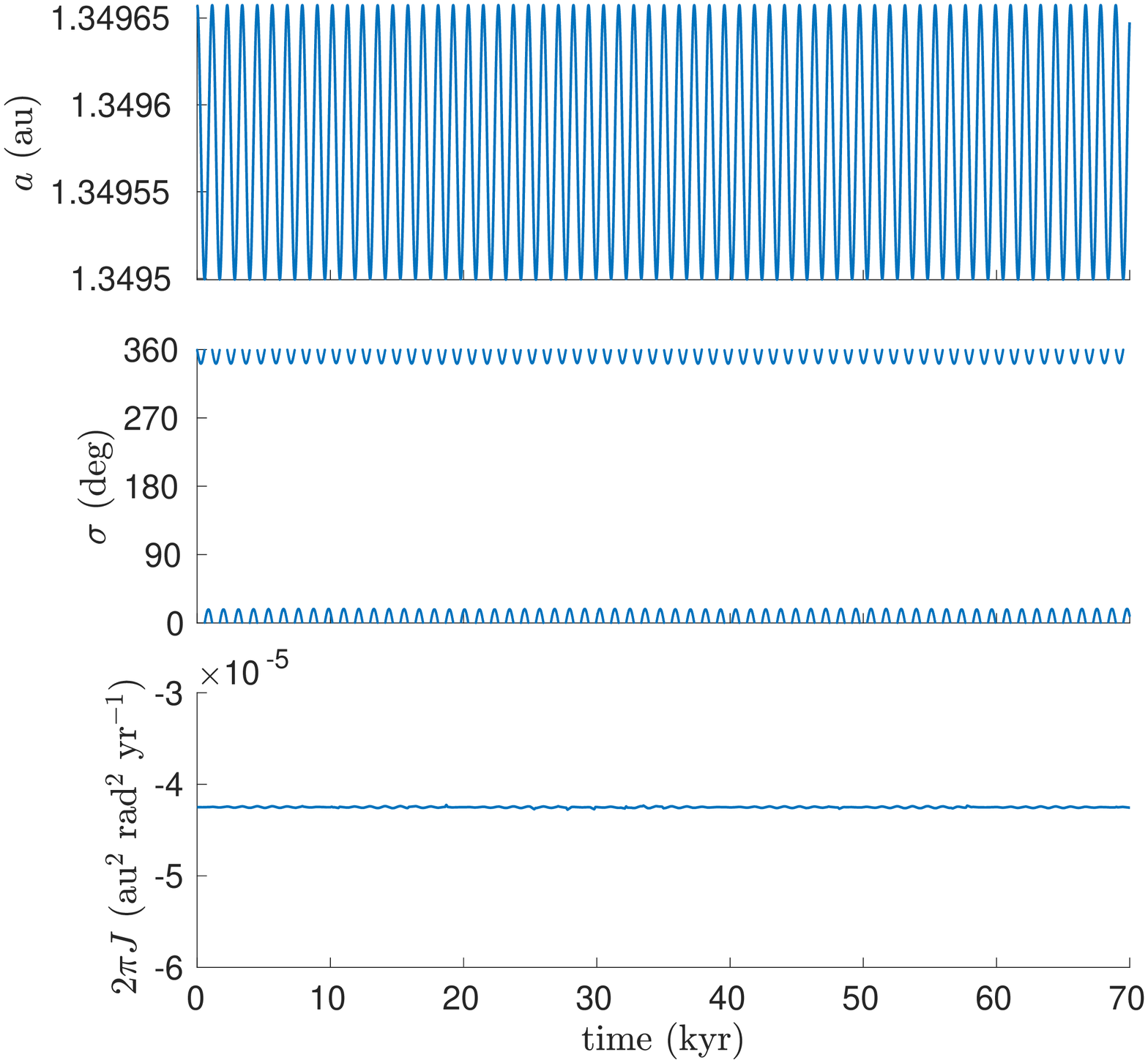}
   \includegraphics[width=0.48\textwidth]{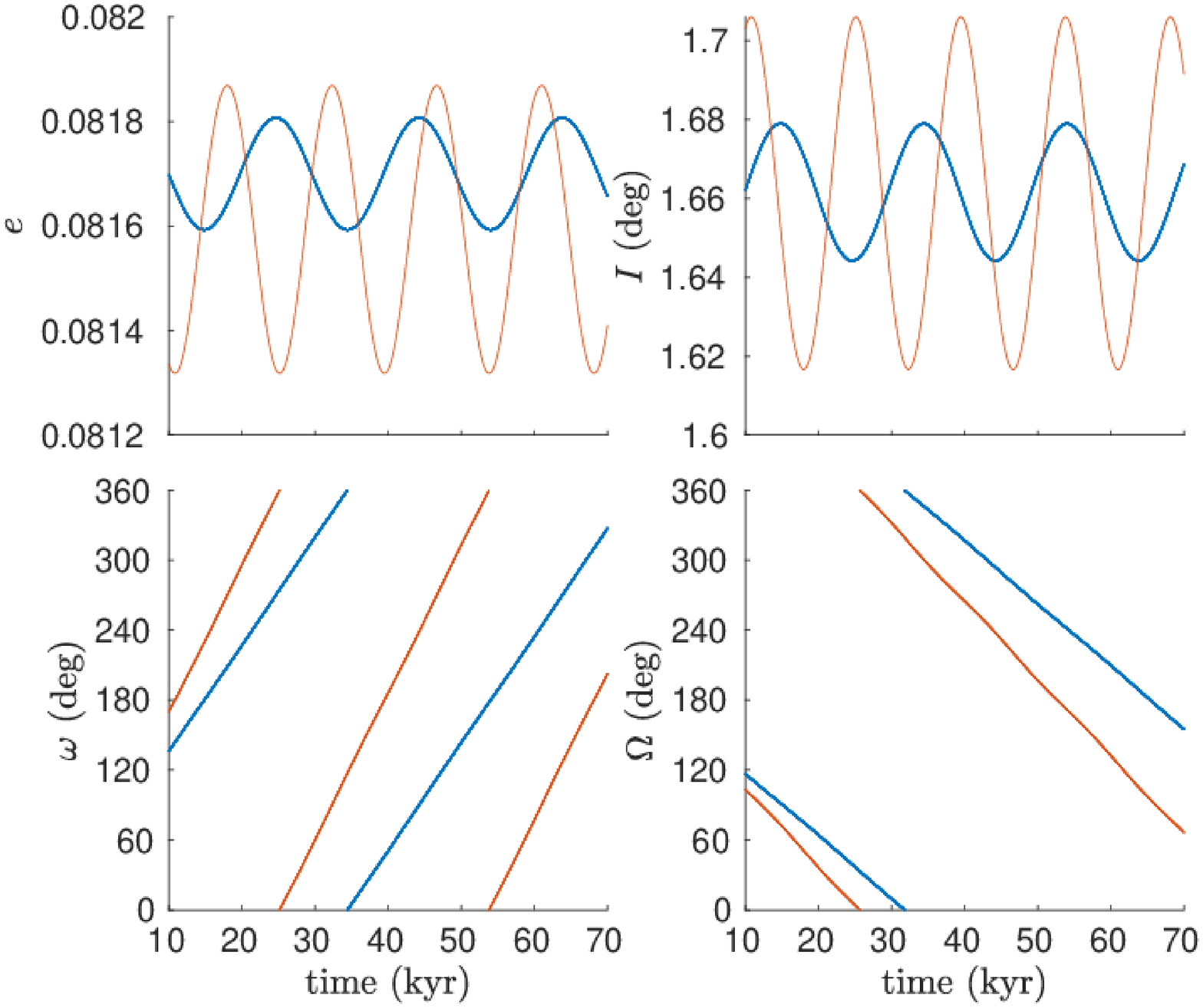}
   \caption{Same as Fig.~\ref{fig:159560_adiabatic}, for asteroid (138911) 2001 AE2.}
   \label{fig:138911_adiabatic}
\end{figure}

We consider here the example of (138911) 2001 AE2, an asteroid in 6:5 mean motion resonance with
Mars that does not cross any planet, and stays in the resonance for the whole integration
timespan of 200 ky (see Fig.~\ref{fig:138911_adiabatic}).
From the numerical integrations we found that the period of the guiding trajectory is
about $T_\sigma \approx 1130$, while the period of circulation of $\omega$ is $T_u \approx
50000$ yr, that results in a ratio of $\xi \approx 0.0226$.  The adiabatic invariant $2\pi
J$ is close to zero, meaning that the object is deep inside the resonance, and its value
is well conserved (see Fig.~\ref{fig:138911_adiabatic}, bottom left panel). The maximum
relative variation with respect to the initial value is smaller than 1\%.

The proper elements reported by NEODyS are $(e_{\min}, e_{\max}) = (0.0813, 0.0819)$,
$(I_{\min}, I_{\max}) = (1.616^\circ, 1.706^\circ)$, and the proper frequencies are $g-s =
45.227$ arcsec yr$^{-1}$, $s = -23.913$ arcsec yr$^{-1}$. 
Figure~\ref{fig:138911_adiabatic} shows the comparisons between the evolution 
of $e, I, \omega, \Omega$ obtained with the non-resonant secular model by 
\citet{gronchi-milani_2001}, and the one obtained with the resonant semi-secular 
model.
It is evident that the resonance significantly affects the secular evolution. 
The oscillation amplitudes of $e$ and $I$ are smaller when the resonance is taken into
account and, more importantly, the proper frequencies $g-s$ and $s$ are significantly different, 
as we can also see from the evolutions of $\omega$ and $\Omega$ shown in
Fig.~\ref{fig:138911_adiabatic}.
Note also that only the evolution between 10 kyr and 70 kyr is shown in these panels,
because of the filtering of periodic oscillations with period smaller than 10 kyr. 

\begin{figure}[!ht]
   \centering
   \includegraphics[width=0.48\textwidth]{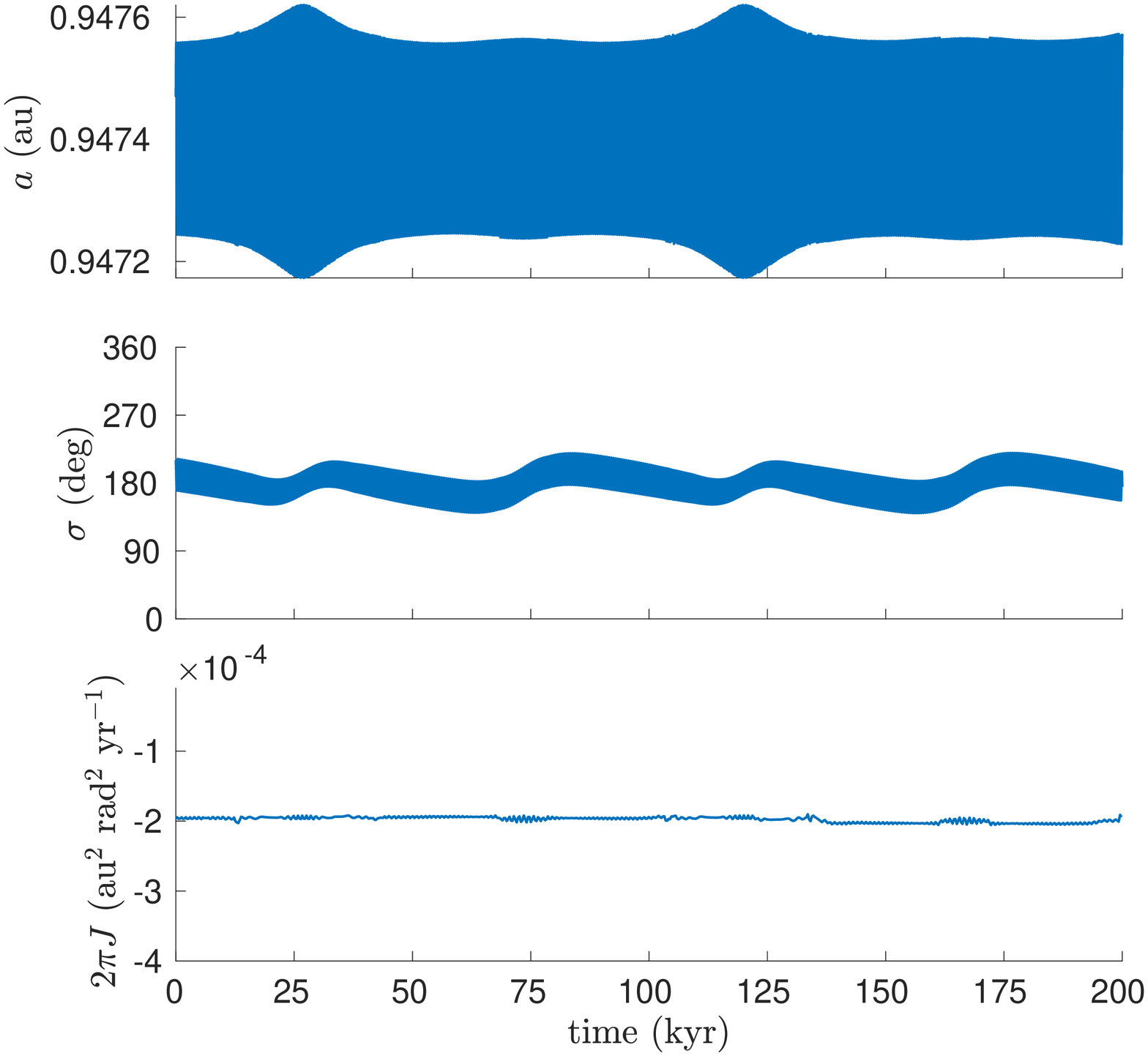}
   \includegraphics[width=0.48\textwidth]{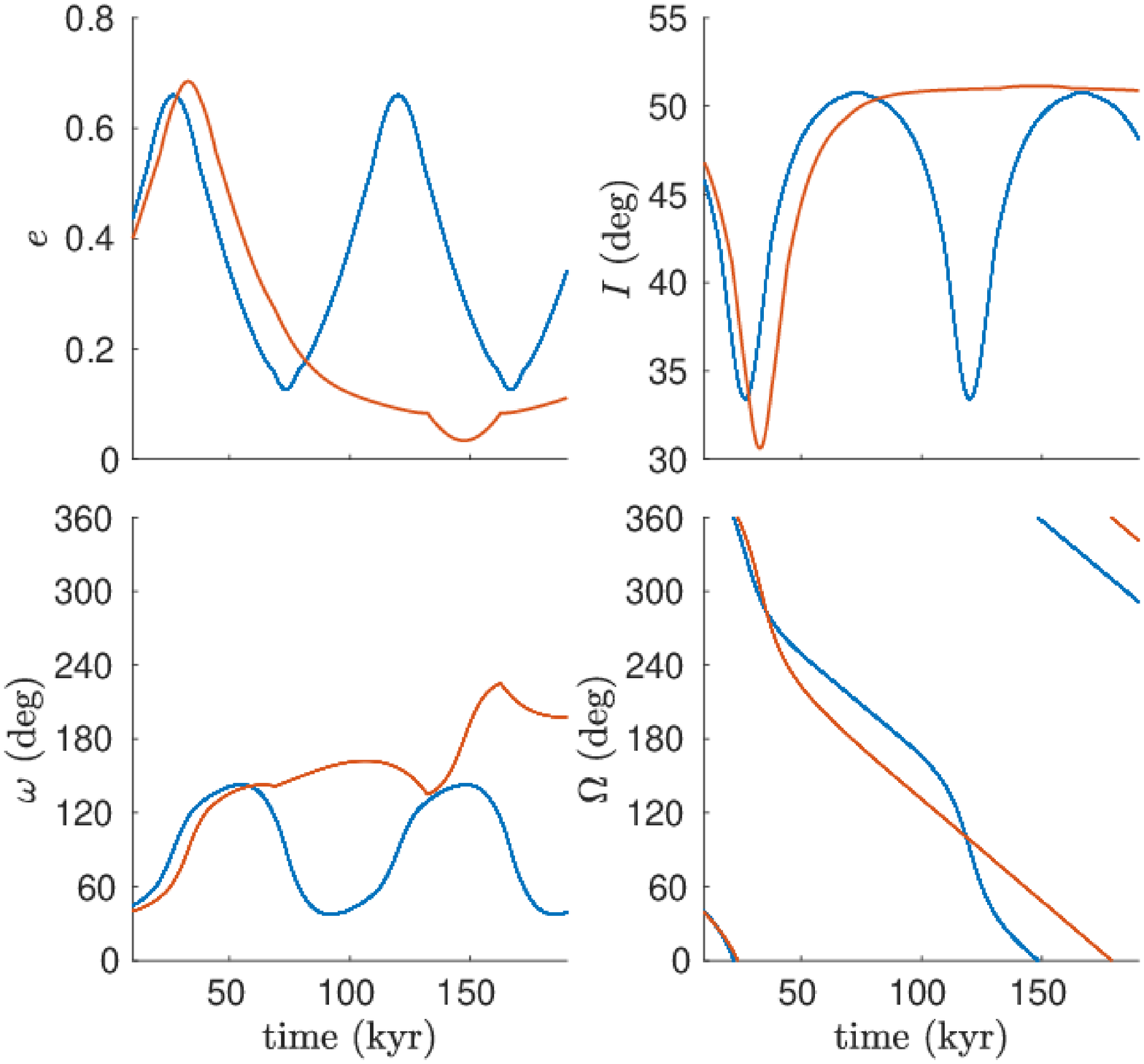}
   \caption{Same as Fig.~\ref{fig:159560_adiabatic}, for asteroid (5381) Sekhmet.}
   \label{fig:5381_adiabatic}
\end{figure}

Another significant example is (5381) Sekhmet, a NEO in 2:3 mean motion resonance with
Venus, that crosses the orbits of Venus itself and that of the Earth. This asteroid stays
in the resonance for the whole integration timespan, while the center of libration of
$\sigma$ and the oscillation amplitude of the semi-major axis $a$ oscillate (see
Fig.~\ref{fig:5381_adiabatic}). The period of the guiding trajectory is about $T_\sigma
\approx 470$ yr, while $\omega$ librates with a period of $T_u \approx 93000$ yr, that results
in a ratio of $\xi \approx 0.005$. 
The evolution of the adiabatic invariant $2\pi J$ is shown in
Fig.~\ref{fig:5381_adiabatic}, bottom left panel, and the maximum relative variation with
respect to the initial value resulted to be about 5\%.

The proper elements reported in NEODyS are $(e_{\min}, e_{\max}) = (0.0338,0.6849)$,
$(I_{\min}, I_{\max}) = (30.619^\circ, 51.142^\circ)$, and the proper frequencies are $g-s =2.821
$ arcsec yr$^{-1}$, $s =-7.914$ arcsec yr$^{-1}$. Figure~\ref{fig:5381_adiabatic}
shows the comparisons of the evolution between the non-resonant secular model, and the filtered
resonant semi-secular model. The oscillations of eccentricity and
inclination are slightly smaller when the resonance is taken into account than when
it is not.
The period of the longitude of
the node $\Omega$ is longer in the non-resonant model and, more importantly, the argument 
of the pericenter $\omega$ librates if we take into account the effects of the resonance,
while it circulates in the non-resonant model. Other objects from Table~\ref{tab:NEOlist}
showing this behavior are (26166), and 2005 YC. 

These two examples already show that taking into account the effect of the resonance is fundamental to
correctly compute the secular evolution, and consequently for the computation of
appropriate proper elements. For the cases of Table~\ref{tab:NEOlist}, we saw that this is 
especially true for resonances with Jupiter, but 
resonances with Venus, the Earth, and Mars can also be important enough to significantly change the
secular evolution.

\subsection{Case B1}
\begin{figure}[!ht]
   \centering
   \includegraphics[width=0.48\textwidth]{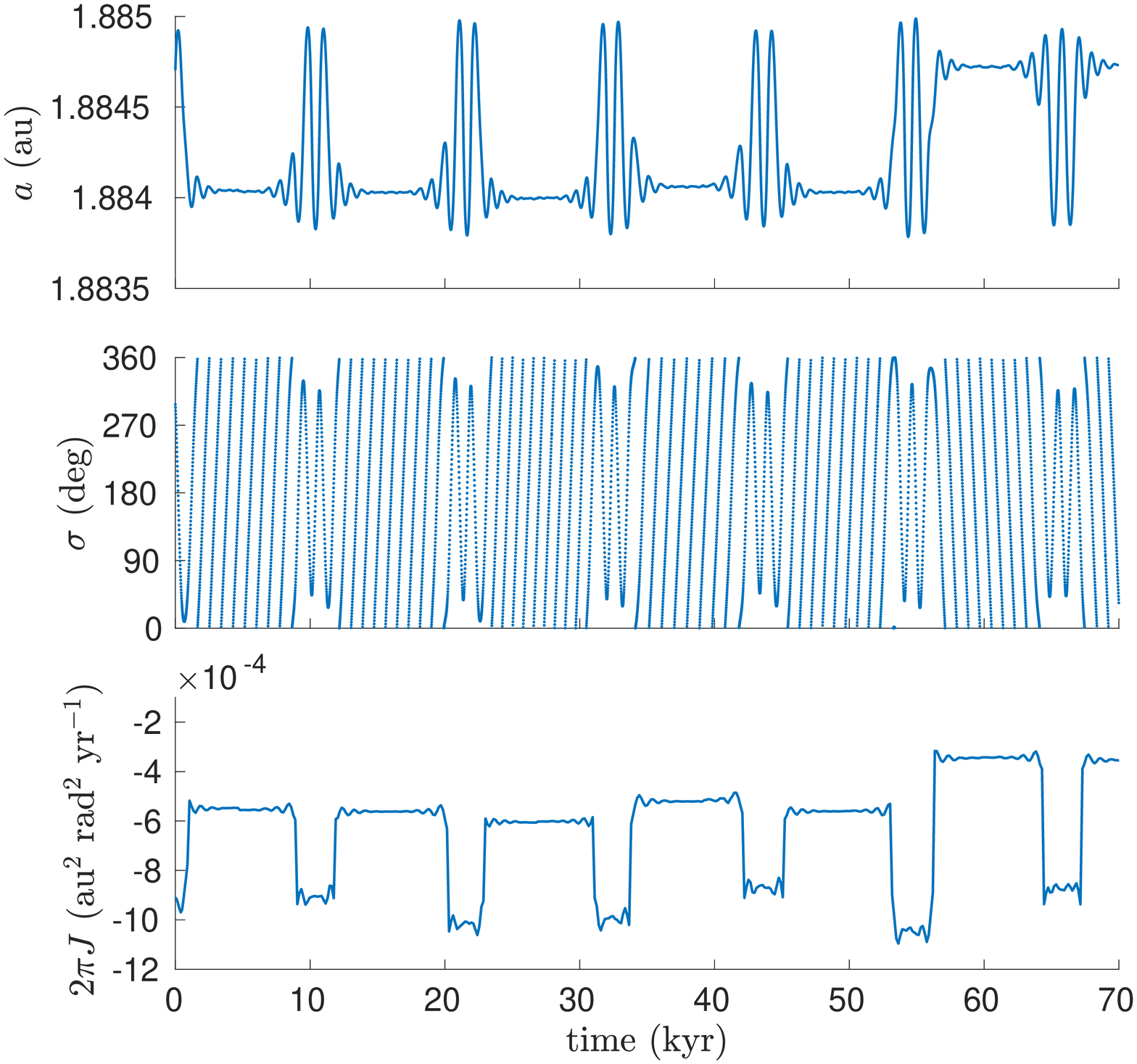}
   \includegraphics[width=0.48\textwidth]{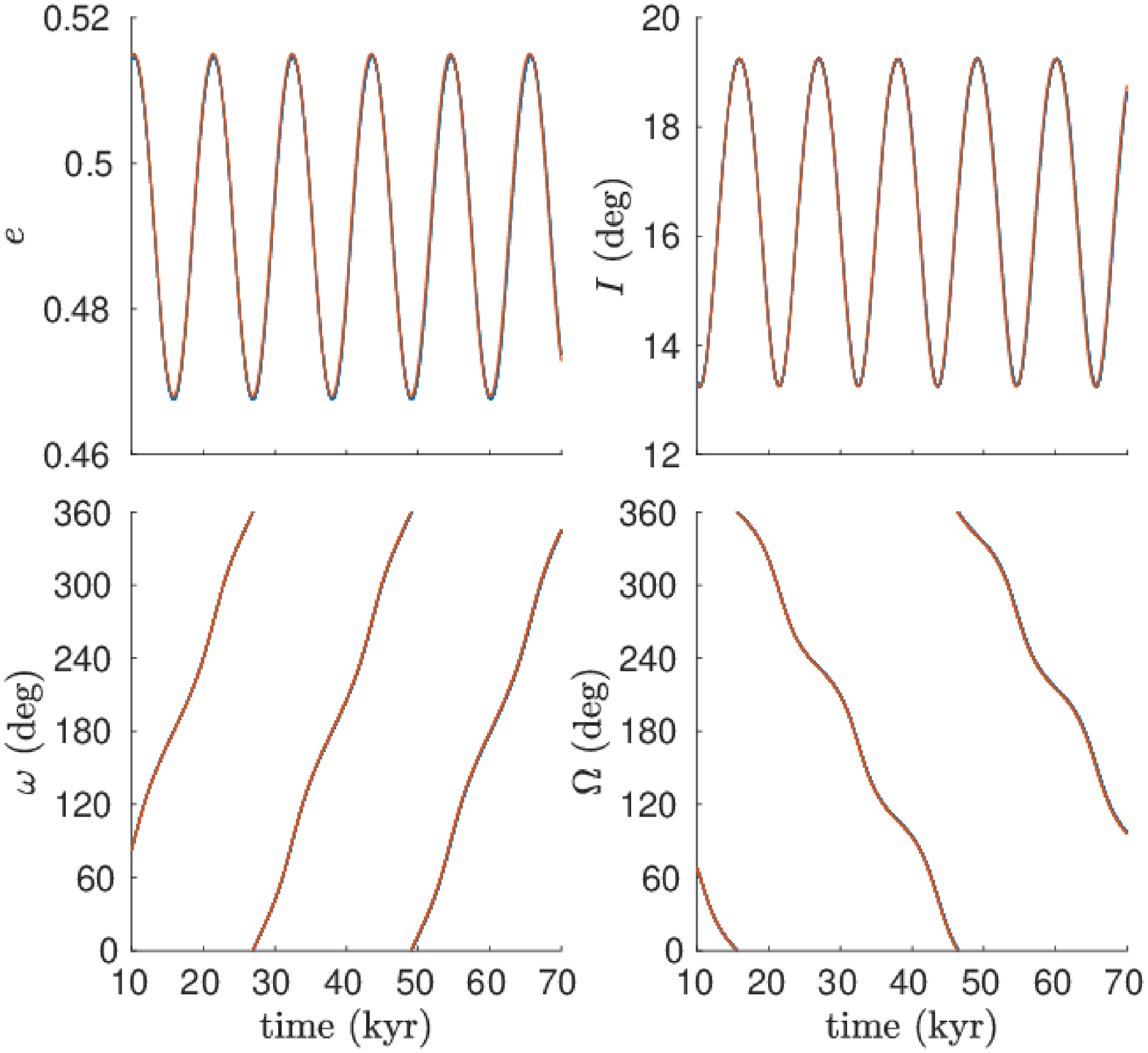}
   \caption{Same as Fig.~\ref{fig:159560_adiabatic}, for asteroid (10636) 1998 QK56.}
   \label{fig:10636_evol}
\end{figure}

We discuss the case of (10636) 1998 QK56, a NEO in 8:11 mean motion resonance with
Mars that crosses the orbit of Mars itself. 
Figure~\ref{fig:10636_evol} shows the evolution of 
$a$ and $\sigma$ for 70 kyr, and we can clearly see the 
switching between libration and circulation happening periodically. The period of the
librating guiding trajectory is $T_\sigma \approx 800$ yr, while the period of the
circulating one is $T_\sigma \approx 1200$ yr. On the other hand, the circulation period of $\omega$ is 
about $T_u \approx 22000$ yr, that results in a ratio $\xi \approx 0.036$ for the libration case
and $\xi \approx 0.054$ for the circulation case. 

\begin{figure}[!ht]
   \centering
   \includegraphics[width=\textwidth]{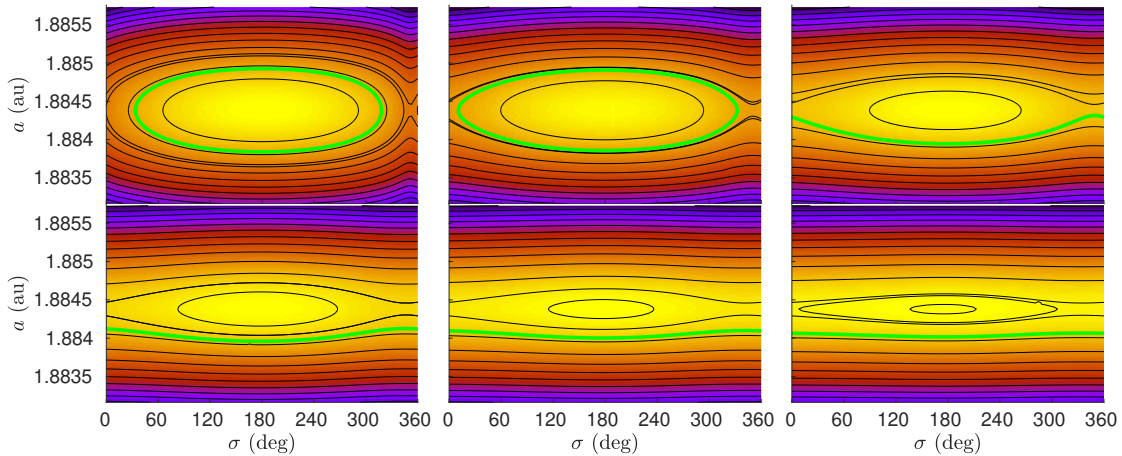}
   \caption{Level curves of the semi-secular Hamiltonian for asteroid (10636) 1998 QK56 in the plane $(a, \sigma)$,
   at times $t = 0, 500, 1000$ years (from the left to the right, top row) and
   $t = 1500, 2000, 2500$ years (from the left to the right, bottom row). The green level
   curve corresponds to the guiding trajectory of the asteroid. See
   Fig.~\ref{fig:10636_evol} for the long-term evolution of all the elements.
}
   \label{fig:10636_lc}
\end{figure}

While the couple $(U,u)$ evolves, the guiding trajectory evolves accordingly, maintaining
roughly constant the value of the area in the $(\Sigma, \sigma)$-plane. Figure~\ref{fig:10636_lc} shows the level curves
of the semi-secular Hamiltonian and the guiding trajectories, at different times. 
During the evolution, the area enclosed by the separatrix curve becomes smaller and
smaller, and this has the effect of pushing the guiding trajectory towards
the separatrix.
At some point of the evolution (e.g. between $t=500$ yr and $t=1000$ yr for the
case shown in Fig.~\ref{fig:10636_lc}), the area enclosed by the separatrix becomes
smaller than the adiabatic invariant $2\pi J$, and a libration motion is not possible
anymore. Thus the guiding trajectory is forced to cross the separatrix, and the
definition (and therefore the value) of the adiabatic invariant changes.
The bottom left panel of Fig.~\ref{fig:10636_evol} shows the evolution of 
$2\pi J$, where we can notice the jumps at times corresponding to the separatrix
crossings. Near the transitions the value of $2\pi J$ substantially drifts because the
dynamics is not adiabatic.
We can also notice that the value of $2\pi J$ is not exactly restored after each crossing,
which may introduce chaos in the semi-secular evolution \citep{wisdom_1985}. 
Separatrix crossings happen very fast if compared to the evolution timescale of $e$ and
$I$, however chaos still produces a long-term diffusion. 

Figure~\ref{fig:10636_evol} shows the comparison between the evolution of the
orbital elements computed with the non-resonant secular model, and the resonant
semi-secular model with digital filtering, and it can be noted that they are essentially
the same.
This shows that the long-term diffusion produced by the separatrix crossings in this example 
is slow enough to be undiscernible over a few tens of thousand years.
The proper elements reported by NEODyS for this NEO are 
$(e_{\min}, e_{\max}) = (0.4681, 0.5152)$, $(I_{\min}, I_{\max}) = (13.249^{\circ},
19.256^\circ)$, and the proper frequencies are $g-s = 58.531$ arcsec yr$^{-1}$,
$s = -40.926$ arcsec yr$^{-1}$, that are very close to the values reported in
Table~\ref{tab:NEOlist}.

Other objects of Table~\ref{tab:NEOlist} that repeatedly switch between libration and circulation and for which we
were able to compute the proper elements are: (163412), (208565), and (309728). In all
these cases, the evolution is the same when taking the resonance into account or not.

\subsection{Case B2}
We consider the NEO (469219) Kamo'oalewa, which is in 1:1 mean motion
resonance with the Earth. The astronomical community gave a lot of attention to this object
upon its discovery, because it is currently in a quasi-satellite configuration with the Earth, meaning
that the critical angle $\sigma$ librates around zero, and therefore it is easily
accessible for space missions \citep{venigalla-etal_2019}. Completely numerical studies of
both the short \citep{delafuente-delafuente_2016} and the long-term dynamics \citep{fenucci-novakovic_2021} 
have been performed.

\begin{figure}[!ht]
   \centering
   \includegraphics[width=0.6\textwidth]{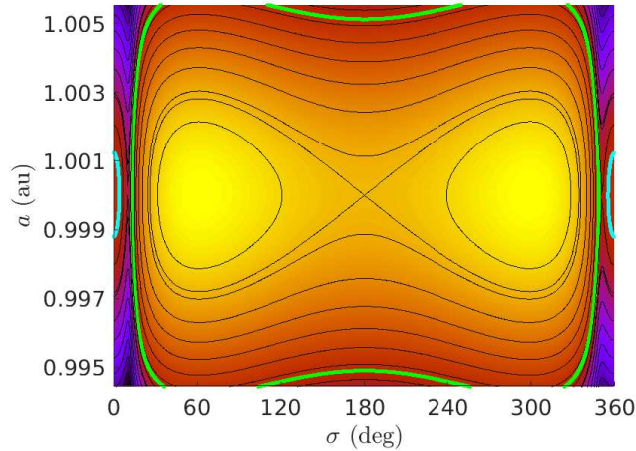}
   \caption{Level curves of the semi-secular Hamiltonian $\mathcal{K}$ in the plane $(a, \sigma)$ for
   asteroid (469219) Kamo'oalewa, at the initial epoch. 
   The level curves corresponding to the value of $\mathcal{K}$ at the
   initial conditions of Kamo'oalewa are highlighted in cyan and green.}
   \label{fig:kamooalewa_LC}
\end{figure}

When the resonance coefficient $h_p$ of the planet is equal to $1$, it may happen that
two resonance islands in the plane $(\Sigma, \sigma)$ appear \citep[see][]{gallardo_2006}, and
this can make the secular dynamics complicated. As the secular variables evolve, a
resonance island can disappear, or a horseshoe-type orbit can appear as well. 
Figure~\ref{fig:kamooalewa_LC} shows the level curves of the semi-secular Hamiltonian
at the initial time. We can see two resonance islands centered at $\sigma = 60^\circ, \
300^\circ$ enclosed by a eight-shaped separatrix curve,
a horseshoe-type orbit surrounding the separatrix, and another small resonance island near
$\sigma = 0^\circ$. The level corresponding to the initial conditions of (469219)
Kamo'oalewa has two different closed curves, colored in green and cyan in
Fig.~\ref{fig:kamooalewa_LC}. Currently, Kamo'oalewa is placed on the cyan level, hence it is librating around $0^\circ$. 

\begin{figure}[!ht]
   \centering
   \includegraphics[width=0.9\textwidth]{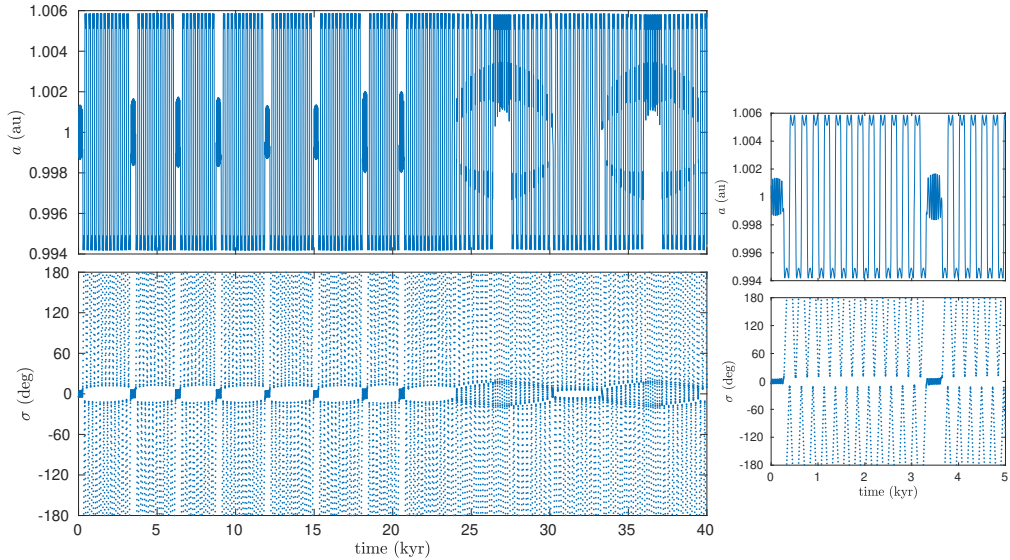}
   \caption{Evolution of semi-major axis (top panel) and critical argument (bottom panel)
   of (469219) Kamo'oalewa given by the semi-secular Hamiltonian, in the time span [0, 40]
   kyr. The panels on the right show a magnification of the evolution in the time interval
   [0, 5] kyr.}
   \label{fig:kamooalewa_a_sigma}
\end{figure}

Figure~\ref{fig:kamooalewa_a_sigma} shows the evolution of the semi-major axis $a$ and of
the critical angle $\sigma$, for the first 40 kyr of propagation with the semi-secular
model. After few thousands years, the orbit passes from libration around $0^\circ$ to a libration around $180^\circ$ 
with a large oscillation amplitude, i.e. a horseshoe-type orbit (as in the
green level curve of Fig.~\ref{fig:kamooalewa_LC}). 
This means that during the evolution of the secular variables, the small 
resonance island around $0^\circ$ that we see in Fig.~\ref{fig:kamooalewa_LC} 
either disappears, or it becomes too small to contain the area that corresponds to the
value of the adiabatic invariant.
While the secular variables continue evolving, the horseshoe orbit slightly shrinks at first, but then it
enlarges again. The small resonance island around $0^\circ$ appears again, and the
horseshoe level curve continues to enlarge until it arrives at the separatrix curve
between horseshoe and quasi-satellite. The separatrix is therefore crossed, and the object is
suddenly placed on a quasi-satellite configuration again. The switching between
quasi-satellite and horseshoe-type happens several times for about 20 kyr of evolution. After
this time, (469219) is placed on a horseshoe-type orbit that opens at about 24 kyr, and the
object passes to a circulation motion, yet another episode of libration occurs
   between 30 and 33 kyr, approximately.

\begin{figure}[!ht]
   \centering
   \includegraphics[width=0.48\textwidth]{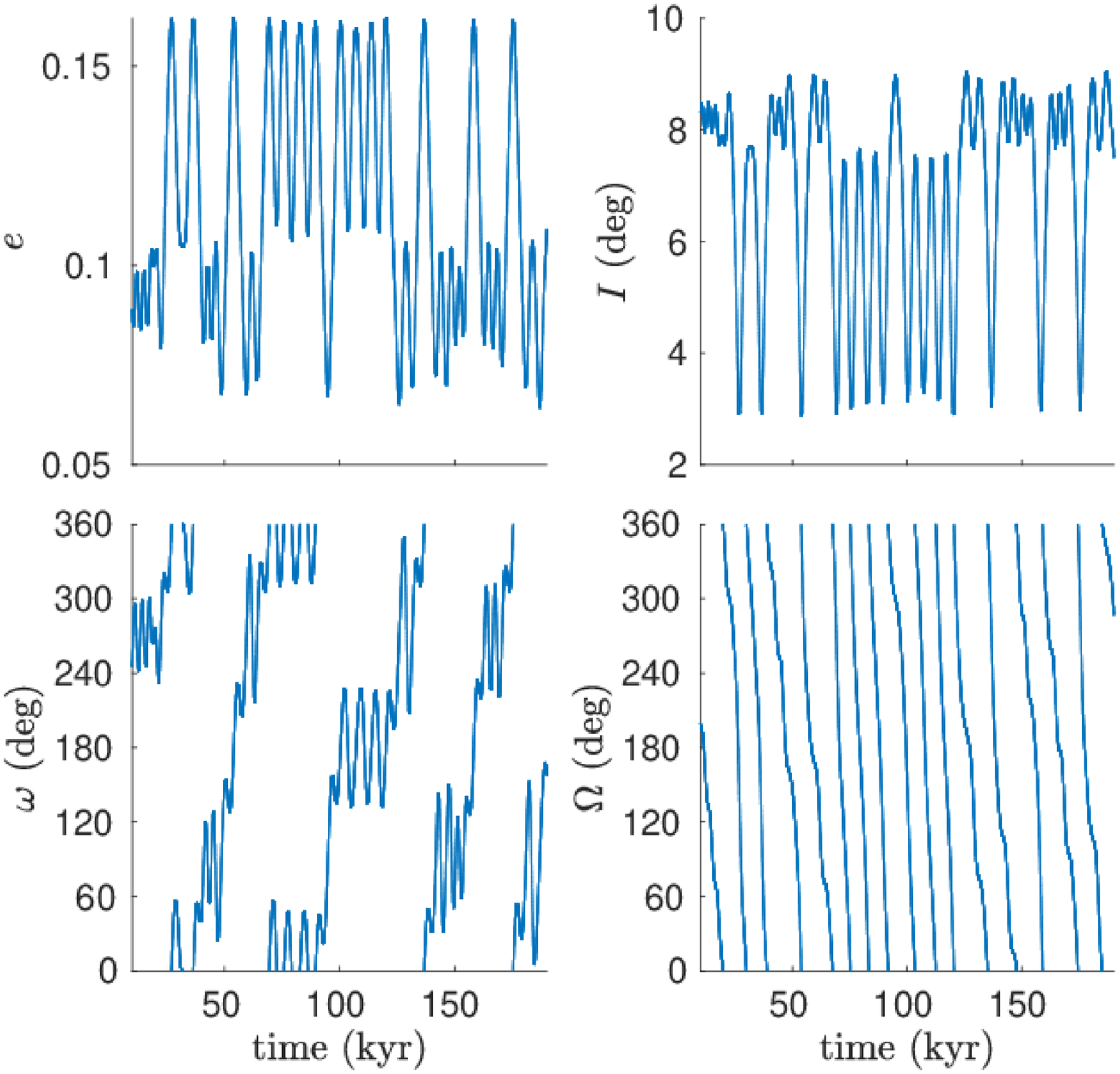}
   \includegraphics[width=0.48\textwidth]{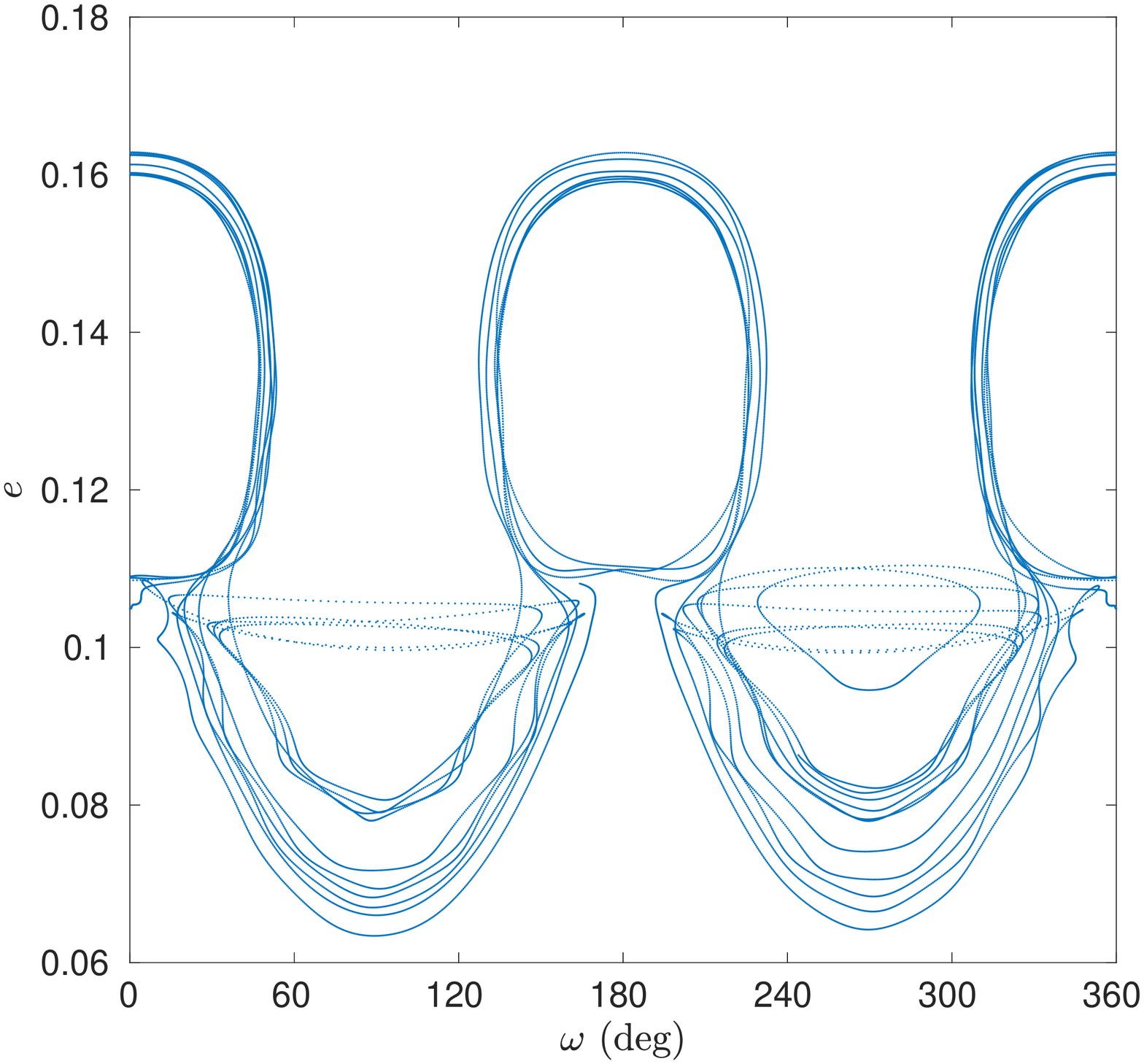}
   \caption{Evolution of $e,I,\omega$ and $\Omega$ (469219) Kamo'oalewa,
   computed with the semi-secular Hamiltonian model and digital filtering.
   The right panel shows the trajectory in the $(\omega, e)$-plane.}
   \label{fig:kamooalewa_w_e}
\end{figure}

The filtered evolution of $e,I,\omega$ and $\Omega$ is reported in
Fig.~\ref{fig:kamooalewa_w_e}, left and central columns. 
Moreover, the right panel of Fig.~\ref{fig:kamooalewa_w_e} shows the trajectory in the $(\omega,e)$-plane. At the
beginning, $\omega$ librates around $270^\circ$ with eccentricity lower than 0.1. 
At about 25 kyr of evolution the argument of perihelion starts circulating, and then it
is trapped again in a libration motion around $90^\circ$, still at low eccentricity values. 
After that, we can notice that $\omega$ spends some time librating around $0^\circ$ or
$180^\circ$, with eccentricity larger than 0.1. However, the switch between 
libration islands in the $(\omega, e)$-plane does not occur regularly. 
It is also worth noting that $\Omega$ changes its slope frequently.

For these types of motions we can not define proper elements
because their secular evolution is chaotic.
It is important to remark that chaos is produced by the appearance and/or disappearance of
resonant islands, and by separatrix crossings that cause a chaotic evolution of the
adiabatic invariant $2\pi J$.
This was already noticed in \citet{namouni_1999, sidorenko-etal_2014}, where the authors studied the general dynamical
structure of the 1:1 mean motion resonance in the restricted three-body problem.

If we intend to apply our method to a large set of resonant NEOs, we need to develop an automatic
detection of such chaotic orbits, or to provide proper elements that are valid on a
shorter timespan (less than 20 kyr for the case of Kamo'oalewa).

\section{Conclusions}
\label{s:conclusions}
In this paper, we described an algorithm for the computation of proper elements of
resonant NEOs that cross the orbit of a planet. In this respect, this work provides an
extension of the \citet{gronchi-milani_2001} approach, where the authors computed proper elements of
NEOs with the assumption that no mean motion resonances nor planetary close approaches occur.
In our model, short periodic perturbations are removed by averaging the Hamiltonian over the fast angles, 
while keeping the resonant argument among the variables. The dynamics of resonant NEOs is
propagated for 200 kyr in the future using the semi-averaged model, and a frequency analysis is then used for the computation
of proper elements and proper frequencies.
We provided some examples of proper elements for known resonant NEOs, and compared
our results with those obtained using the non-resonant model by
\citet{gronchi-milani_2001}.
For some objects, the mean-motion resonance has no noticeable effect on the dynamics, and
the resonant proper elements that we obtain are similar to the non-resonant ones. For
other objects, on the contrary, the mean-motion resonance strongly alters the long-term
dynamics, and reliable proper elements can be obtained only if the resonance is taken into
account.

In this paper we provided results only for a limited number of NEOs, that we
identified to be in a mean-motion resonance by using pure $N$-body simulations.
However, the method presented here can be applied to the full set of resonant NEOs, to
build a complete database of the resonant proper elements.

\appendix

\section{Crossing singularity}
\label{app:crossing}
\subsection{The minimum orbit intersection distance}
Let $(E, \ell), (E', \ell') \in \R^6$ be two sets of orbital elements of two Keplerian
orbits with a common focus. The components $E, E' \in \R^5$ describe the shape of the orbit, while $\ell,
\ell' \in S^1$ are the mean anomalies. 
We denote with $\calE = (E,E') \in \R^{10}$ the couple of the orbit configurations, and
with $V = (\ell,\ell') \in \T^2 = S^1 \times S^1$ the parameters along the orbits. 
We choose a reference frame centered at the common focus and we denote with $\CHI(E,\ell),
\CHI'(E', \ell')$ the Cartesian coordinates of the two bodies. 
For a given configuration $\calE$, we define the Keplerian distance function
$d$ as
\begin{equation}
   d: \T^2 \to \R, \quad d(\calE, V) = |\CHI-\CHI'|.
   \label{eq:kepDistFunction} 
\end{equation}
Let $V_h=V_h(\calE)$ be a local minimum point\footnote{Here the subscript $h$ is used to
refer to a local minimum point, it has nothing to do with the integer of the resonant
combination of Eq.~\eqref{eq:resonantAngle}.} of the Keplerian distance function and consider the maps
\begin{equation}
   \calE \mapsto d_h(\calE) = d(\calE, V_h), \quad \calE \mapsto d_{\text{min}}(\calE) =
   \min_{h} d_h(\calE, V_h).
\end{equation}
A configuration $\calE$ is non-degenerate if all the critical points of the Keplerian
distance function are non-degenerate. If $\calE$ is non-degenerate, then there exists a
neighborhood $\calW \subseteq \R^{10}$ of $\calE$ such that the maps $d_h$, restricted to $\calW$,
do not have bifurcations. 

The functions $d_h$ and $d_{\text{min}}$ are not smooth at crossing configurations, and their
derivatives do not exist. However, it is possible to define analytical maps in a
neighborhood of a non-degenerate crossing configuration $\mathcal{E}_c$ by choosing an
appropriate sign for the maps.
We summarize the procedure to deal with the crossing singularity of $d_h$, the
procedure for $d_{\text{min}}$ being the same. We consider the points on the two ellipses 
corresponding to the local minimum points $V_h = (\ell_h, \ell'_h)$ of $d^2$, i.e.
\begin{equation}
   \CHI_h = \CHI(E, \ell_h), \quad \CHI'_h = \CHI'(E',\ell'_h).
\end{equation}
We denote with $\tau_h, \tau'_h$ the tangent vectors to the trajectories $E,E'$ at these
points, i.e.
\begin{equation}
   \tau_h = \frac{\partial \CHI}{\partial \ell}(E,\ell_h), \quad
   \tau'_h = \frac{\partial \CHI'}{\partial \ell'}(E',\ell'_h),
\end{equation}
and their cross product 
\begin{equation}
   \tau^*_h = \tau_h \times \tau'_h.
\end{equation}
\noindent We define also $\Delta = \CHI-\CHI', \Delta_h = \CHI_h-\CHI'_h$.
The vector $\Delta_h$ joins the points attaining a local minimum value of $d^2$, hence
$|\Delta_h|=d_h$. 
From the definition of critical points of $d^2$, both vectors $\tau_h,
\tau'_h$ are orthogonal to $\Delta_h$, therefore $\tau^*_h$ and $\Delta_h$ are
parallel. Denoting with $\widehat{\tau}^*_h,\widehat{\Delta}_h$ the corresponding unit
vectors, the distance with sign 
\begin{equation}
   \tilde{d}_h = \big( \widehat{\tau}^*_h \cdot \widehat{\Delta}_h \big) d_h,
   \label{eq:distWithSign}
\end{equation}
is an analytic function in a neighborhood of a crossing configuration, provided that
$\tau_h$ and $\tau'_h$ are not parallel, situation happening only when the trajectories 
are tangent at the crossing point \citep{gronchi-tommei_2007}.
The derivatives of $\tilde{d}_h$ with respect to the component $\calE_k, \, k=1,\dots,10$
of $\calE$ are given by
\begin{equation}
   \frac{\partial \tilde{d}_h}{\partial\calE_k} = \widehat{\tau}^*_h \cdot
   \frac{\partial \Delta}{\partial\calE_k}(\calE,V_h). 
   \label{eq:derDistWithSign}
\end{equation}

\subsection{Extraction of the singularity}
\label{ss:extraction}
Denote by $\calE_c$ a non-degenerate crossing configuration with only one
crossing point. We choose the index $h$ such that
$d_h(\calE_c) = 0$. For each $\calE$ in a neighborhood of $\calE_c$
we consider the Taylor development of $V\mapsto d^2({\cal E},V) =
|\mathcal{X}-\mathcal{X}'|^2$, in a neighborhood
of the local minimum point $V_h=V_h(\calE)$, i.e.
\begin{equation}
d^2({\cal E},V) = d_h^2({\cal E}) + \frac{1}{2}(V-V_h)\cdot
  H_h({\cal E})(V-V_h) + {\cal R}^{(h)}({\cal E},V)\,,
\label{eq:taylor_d2}
\end{equation}
where 
\begin{equation}
H_h({\cal E}) = \frac{\partial^2 d^2}{\partial
  V^2}(\calE,V_h(\calE)),
\end{equation}
is the Hessian matrix of $d^2$ at $V_h=(\ell_h,\ell_h')$, and
${\cal R}^{(h)}$ is the Taylor remainder.
We introduce the approximated distance 
\begin{equation}
\delta_h = \sqrt{d_h^2 + (V-V_h)\cdot {\cal A}_h(V-V_h)}\,,
\label{approxdist}
\end{equation}
where
\begin{equation}
\calA_h = \frac{1}{2}H_h = \left[
\begin{array}{cc}
|\tau_h|^2+ \displaystyle\frac{\partial^2\mathcal{X}}{\partial \ell^2}(E,\ell_h)\cdot\Delta_h  
                   &-\tau_h\cdot\tau_h' \cr
                   &\cr
-\tau_h\cdot\tau_h' &|\tau_h'|^2 -\displaystyle \frac{\partial^2\mathcal{X}'}{\partial \ell'^2}(E',\ell_h')\cdot\Delta_h \cr
\end{array}
\right],
\end{equation}
and
\begin{equation}
\Delta_h = \Delta_h(\calE)\,,
\quad
\tau_h = \frac{\partial {\mathcal X}}{\partial \ell}(E,\ell_h)\,,
\quad
\tau_h' = \frac{\partial {\mathcal X}'}{\partial \ell'}(E',\ell_h')\ .
\end{equation}
If the matrix $\calA_h$ is non-degenerate, then it is positive definite since
$V_h$ is a minimum point, and this property holds in a suitably chosen neighborhood
$\calW$ of $\calE_c$. The matrix $\calA_h$ is degenerate at the crossing
configuration if and only if the tangent vectors $\tau_h,\tau_h'$ are parallel, therefore
in the following we always assume that the crossing is not tangent.

To extract the singularity at an orbit crossing, we split the integral as
   \begin{equation}
   \int_{\T^2}^{}\frac{1}{d}\, \text{d}\ell \text{d}\ell' =
   \int_{\T^2}^{}\bigg(\frac{1}{d}-\frac{1}{\delta_h} \bigg) \text{d}\ell \text{d}\ell' +
   \int_{\T^2}^{}\frac{1}{\delta_h}\text{d}\ell \text{d}\ell'.
      \label{eq:splitHamiltonian}
   \end{equation}
Let us set $\mathcal{S} = \{ \calE \in \calW: \, d_h(\calE)=0 \}$, and denote with
$y_k \in \{ \Sigma, U, V, \sigma, u, v \}$ one of the coordinates. 
The derivatives of the first term in the right-hand side of Eq.~\eqref{eq:splitHamiltonian} are integrable, and the map
\begin{equation}
   \calW \setminus \mathcal{S} \ni \calE \mapsto \int_{\T^2}^{}\frac{\partial}{\partial
   y_k}\bigg(\frac{1}{d}-\frac{1}{\delta_h}\bigg) \text{d}\ell \text{d}\ell',
   \label{eq:derIntReminder}
\end{equation}
can be extended continuously to the whole set $\calW$.
To compute the derivatives in Eq.~\eqref{eq:derIntReminder} we can use
\begin{equation}
   \frac{\partial}{\partial y_k}\bigg(\frac{1}{\delta_h}\bigg) =
   -\frac{1}{2\delta_h^3}\frac{\partial \delta_h^2}{\partial y_k}.
   \label{eq:d1sudelta}
\end{equation}
From Eq.~\eqref{eq:taylor_d2} we obtain the derivatives of the approximated distance as
\begin{equation}
   \frac{\partial \delta_h^2}{\partial y_k} = \frac{\partial d_h^2}{\partial y_k} - 2
   \frac{\partial V_h}{\partial y_k}\cdot \calA_h(V-V_h) +
   (V-V_h)\cdot\frac{\partial \calA_h}{\partial y_k}(V-V_h).
   \label{eq:derDeltah2}
\end{equation}
The derivatives of $V_h$ are computed by differentiating the relation 
\begin{equation}
   \frac{\partial}{\partial y_k}d_h^2(\calE, V_h(\calE)) = 0,
\end{equation}
which holds since $(\calE, V_h(\calE))$ is a stationary point of $d^2$. Hence
\begin{equation}
   \frac{\partial V_h}{\partial y_k}(\calE) = -
   [H_h(\calE)]^{-1}\frac{\partial}{\partial y_k} \nabla_V
   d^2(\calE,V_h(\calE)).
   \label{eq:derVh}
\end{equation}
%
Note that the derivatives of $1/d$ are obtained with standard computations, and
they can be expressed through the derivatives of the position of the asteroid with respect to
the Delaunay variables.
On the other hand, the average over $\T^2$ of the derivatives of $1/\delta_h$ are
non-convergent integrals for $\calE \in \mathcal{S}$, and contains the main 
part of the singularity of the vector field.

\subsection{Integration of $1/\delta_h$ and its derivatives}
Let $(\calE_c, V_h(\calE_c))$ be a crossing configuration. We consider the transformations
\begin{equation}
   {\cal T}_h(V) = V + V_h, \quad {\cal L}_h(V) = \sqrt{{\cal A}_h}\ V,
   \label{eq:twoTransforms}
\end{equation}
where $\sqrt{\mathcal{A}_h}$ is defined as the unique positive definite matrix such that
$(\sqrt{\mathcal{A}_h})^2 = {\cal A}_h$. With these constraints, the entries $a_{ij}$ of
$\sqrt{\mathcal{A}_h}$ are
\begin{equation}
  a_{11}  = \frac{\alpha+A_{11}}{\sqrt{2\alpha + A_{11}+A_{22}}}, \quad
  a_{22}  =  \frac{\alpha+A_{22}}{\sqrt{2\alpha + A_{11}+A_{22}}}, \quad
  a_{12}  = \frac{A_{12}}{\sqrt{2\alpha + A_{11}+A_{22}}},
   \label{eq:aij}
\end{equation}
where $\alpha = \sqrt{\det \mathcal{A}_h}$, and $A_{ij}$ are the entries of
$\mathcal{A}_h$.
Using these transformations to change the coordinates in the integral, we get 
\begin{equation}
   \int_{\T^2}^{}\frac{1}{\delta_h}\text{d}\ell \text{d}\ell' = \int_{{\cal T}_h(\mathbb{T}^2)}\frac{1}{\delta_h} \text{d}V =
   \frac{1}{{\sqrt{\det\A_h}}}
\int_{{\cal L}_h(\mathbb{T}^2)}\frac{1}{\sqrt{d_h^2 + |W|^2}}\text{d}W
\label{eq:int1sudelta}
\end{equation}
where
\begin{equation}
W = {\cal L}_h\circ{\cal T}_h^{-1}(V) = {\cal L}_h(V - V_h).
\end{equation}
Let us consider the points $P_1 \equiv (\pi,\pi), P_2 \equiv (-\pi,\pi), P_3 \equiv (-\pi,-\pi), 
P_4 \equiv (\pi,-\pi)$
and their images $Q_j \equiv (x_j,y_j), j=1,\dots,4$ through ${\cal L}_h$, so that
\begin{equation}
(x_1, y_1) = \pi (a_{11}+a_{12}, a_{12}+a_{22}), \quad
(x_2, y_2) = \pi( -a_{11}+a_{12}, -a_{12}+a_{22}), 
\end{equation}
\begin{equation}
(x_3, y_3) = -(x_1,y_1), \quad
(x_4, y_4) = -(x_2,y_2).
\end{equation}
Set $P_5=P_1$ and, for $j=1,\ldots,4$, let $\mathscr{R}_j$ be the straight line
passing through the points $P_j, P_{j+1}$, i.e.
\begin{equation}
\xi_j(y-y_j) = \eta_j(x-x_j),
\end{equation}
where $\xi_j = x_{j+1}-x_j, \ \eta_j = y_{j+1}-y_j.$
Introducing polar coordinates $(\rho,\theta)$ such that
$W = (\rho\cos\theta, \rho\sin\theta)$,
we can write these lines in polar form
\begin{equation}
\mathscr{R}_j =
\bigl\{\bigl(r_j(\theta)\cos\theta,r_j(\theta)\sin(\theta)\bigr): \theta\in(\bar{\theta}_j,\bar{\theta}_j+\pi)\bigr\}
\end{equation}
with
\begin{equation}
r_j(\theta) = \frac{\xi_j y_j -
  \eta_jx_j}{\xi_j\sin\theta-\eta_j\cos\theta}
\end{equation}
and
\begin{equation}
\bar{\theta}_j = \left\{
\begin{array}{ll}
  \arctan({\eta_j}/{\xi_j}), \qquad&\xi_j\neq 0,\cr
  \pi/2,                           &\xi_j= 0.\cr
\end{array}
\right.
\end{equation}
Note that
\begin{equation}
(\xi_1,\eta_1) = -2\pi(a_{11}, a_{12}),\qquad
(\xi_2,\eta_2) = -2\pi(a_{12}, a_{22}),
\end{equation}
\begin{equation}
(\xi_3,\eta_3) = -(\xi_1,\eta_1), \qquad
(\xi_4,\eta_4) = -(\xi_2,\eta_2),
\end{equation}
so that, for each $j=1,\ldots,4$,
\begin{equation}
\xi_j y_j - \eta_j x_j = -2\pi^2\sqrt{\det{\cal A}_h}
\end{equation}
and
\begin{equation}
r_1(\theta) = \frac{\pi\sqrt{\det{\cal A}_h}
}{a_{11}\sin\theta-a_{12}\cos\theta},
\qquad
r_2(\theta) = \frac{\pi\sqrt{\det{\cal A}_h}}{a_{12}\sin\theta-a_{22}\cos\theta},
\end{equation}
\begin{equation}
r_3(\theta) = -r_1(\theta), \qquad r_4(\theta) = -r_2(\theta). 
\end{equation}
With these changes of coordinates, Eq.~\eqref{eq:int1sudelta} becomes
\begin{equation}
   \int_{{\cal T}_h(\mathbb{T}^2)}\frac{1}{\delta_h}\text{d}\ell \text{d}\ell' = \frac{1}{\sqrt{\det {\cal A}_h}}\biggl(
   \sum_{j=1}^4\int_{\theta_j}^{\theta_{j+1}}\sqrt{d_h^2 + r_j^2(\theta)}\text{d}\theta - 2\pi d_h\biggr)
   \label{eq:int1sudeltah}
\end{equation}
with
\begin{equation}
\cos\theta_j = \frac{x_j}{\sqrt{x_j^2+y_j^2}}, \qquad \sin\theta_j =
\frac{y_j}{\sqrt{x_j^2+y_j^2}},
\end{equation}
and
\begin{equation}
\theta_1<\theta_2<\theta_3<\theta_4<\theta_5 = 2\pi+\theta_1.
\end{equation}
The integrals in Eq.~\eqref{eq:int1sudeltah} are bounded, hence they are
differentiable functions of the elements. On the contrary, the term $-2\pi
d_h/\sqrt{\det{\cal A}_h}$ is not differentiable at $\calE = \calE_c\in\mathcal{S}$, and the
loss of regularity is due only to this term.
The derivatives of Eq.~\eqref{eq:int1sudeltah} with respect to $y_k \in \{ \Sigma, U, V,
\sigma, u,v \}$ can be computed by exchanging the integral sign
and the derivative, i.e.
\begin{equation}
\begin{split}
\frac{\partial}{\partial y_k}\int_{\mathbb{T}^2}\frac{1}{\delta_h}\text{d}\ell \text{d}\ell' &=
\Bigl(\frac{\partial}{\partial y_k}\frac{1}{\sqrt{\det {\cal A}_h}}\Bigr)
\biggl(\sum_{j=1}^4\int_{\theta_j}^{\theta_{j+1}}\sqrt{d_h^2 + r_j^2(\theta)}\text{d}\theta - 2\pi d_h\biggr)\\
&+ \frac{1}{\sqrt{\det{\cal
      A}_h}}\biggl(\sum_{j=1}^4\int_{\theta_j}^{\theta_{j+1}}
\frac{d_h\frac{\partial d_h}{\partial y_k}+ r_j(\theta)\frac{\partial
    r_j}{\partial y_k}(\theta) }{\sqrt{d_h^2 + r_j^2(\theta)}}\text{d}\theta
- 2\pi \frac{\partial d_h}{\partial y_k}\biggr).
\end{split}
\label{derivapprox}
\end{equation}
The term $-2\pi d_h/\sqrt{\det{\calA_h}}$ is not differentiable at the orbit
crossing, however the derivatives admit two analytic extensions $(\frac{\partial
\mathcal{K}_{\text{sec}}}{\partial y_k})_h^{\pm}$ on 
$\calW^+ = \calW \cap \{\tilde{d}_h > 0\}$ and $\calW^- = \calW \cap \{\tilde{d}_h < 0\}$,
where ${\cal W}$ is a neighborhood of the crossing configuration ${\cal E}_c$ where $\tilde{d}_h$ is defined, 
and ${\cal A}_h$ is non-degenerate \citep{gronchi-tardioli_2013}.
Moreover, the jump in the derivatives passing from $\mathcal{W}^+$ to $\mathcal{W}^-$ is given by
\begin{equation}
   \begin{split}
      \text{Diff}_h\bigg( \frac{\partial\mathcal{K}_{\text{sec}}}{\partial y_k} \bigg) & := 
      \bigg( \frac{\partial \mathcal{K}_{\text{sec}}}{\partial y_k}\bigg)_h^- - 
      \bigg( \frac{\partial \mathcal{K}_{\text{sec}}}{\partial y_k}\bigg)_h^+ \\
      & = \frac{1}{\pi}\bigg[ \frac{\partial}{\partial y_k}\bigg( \frac{1}{\sqrt{\det
      \calA_h}}\bigg)\tilde{d}_h +
   \frac{1}{\sqrt{\det\calA_h}}\frac{\partial\tilde{d}_h}{\partial y_k}\bigg]. 
   \end{split}
   \label{eq:derJump}
\end{equation}

\section*{Acknowledgments}
We thank the anonymous referee for the comments and suggestions that helped us to improve the manuscript.
MF and GFG have been supported by the H2020 MSCA ETN Stardust-Reloaded, grant agreement
number 813644.
GFG also acknowledges the project MIUR-PRIN 20178CJA2B “New frontiers of Celestial Mechanics:
theory and applications” and the GNFM-INdAM (Gruppo Nazionale per la Fisica Matematica).

\section*{Data availability statement}
The datasets generated during and/or analysed during the current study are available from
the corresponding author on reasonable request.

\bibliography{holybib}{}
\bibliographystyle{apalike85}

\end{document}